\documentclass[runningheads]{llncs}
\usepackage[utf8]{inputenc}

 \usepackage{enumitem}
\usepackage{amsmath}
\usepackage{amssymb}
\usepackage{math_definition}
\usepackage{color}

\usepackage{comment}
\usepackage{algorithm}
\usepackage{algorithmicx}
\usepackage[algo2e,ruled,vlined,linesnumbered]{algorithm2e}
\usepackage{hyperref}
\usepackage{graphicx}
\usepackage{float}
\usepackage{subfigure}
\usepackage{fancyhdr}
\usepackage{pgfplots}
\pgfplotsset{compat=1.15}
\usetikzlibrary{arrows}
\usetikzlibrary[patterns]
\usepackage{pgfplots}
\pgfplotsset{compat=1.15}
\usepackage{mathrsfs}
\usetikzlibrary{arrows}

\usepackage{caption}
\title{Local Search for Solving Satisfiability of Polynomial Formulas}

\author{Haokun Li \and Bican Xia \and Tianqi Zhao}
\institute{School of Mathematical Sciences, Peking University, Beijing, China.  \\\email{\{haokunli, zhaotq\}@pku.edu.cn} \\
\email{xbc@math.pku.edu.cn}
}
\begin{document}
\maketitle

\begin{abstract}
Satisfiability Modulo the Theory of Nonlinear Real Arithmetic, SMT(NRA) for short, concerns the satisfiability of \emph{polynomial formulas}, which are quantifier-free Boolean combinations of polynomial equations and inequalities with integer coefficients and real variables. 
In this paper, we propose a local search algorithm for a special subclass of SMT(NRA), where all constraints are strict inequalities.
An important fact is that, given a polynomial formula with $n$ variables, the zero level set of the polynomials in the formula decomposes the $n$-dimensional real space into finitely many components (cells) and every polynomial has constant sign in each cell.
The key point of our algorithm is a new operation based on real root isolation, called \emph{cell-jump}, which updates the current assignment along a given direction such that the assignment can `jump' from one cell to another.  
One cell-jump may adjust the values of several variables while traditional local search operations, such as \emph{flip} for SAT and \emph{critical move} for SMT(LIA), only change that of one variable.
We also design a two-level operation selection 
to balance the success rate and efficiency.
Furthermore, our algorithm can be easily generalized to a wider subclass of SMT(NRA) where polynomial equations linear with respect to some variable are allowed.  
Experiments show the algorithm is competitive with state-of-the-art SMT solvers, and performs particularly well on those formulas with high-degree polynomials. 
\end{abstract}
\vspace{-2mm}
\section{Introduction}
\vspace{-1mm}
Satisfiability modulo theories (SMT) refers to the problem of determining
whether a first-order formula is satisfiable with respect to (w.r.t.) certain theories, such as the theories of linear integer/real arithmetic, nonlinear integer/real arithmetic and string.
In this paper,
we are concerned the theory of nonlinear real arithmetic (NRA) and
restrict our attention to the problem of solving satisfiability of polynomial formulas.

Solving polynomial constraints has been a central problem in the development of mathematics.
In 1951, Tarski's decision procedure \cite{10.1007/978-3-7091-9459-1_3} made it possible to solve polynomial constraints in an algorithmic way.
However, Tarski's algorithm is impractical because of its super-exponential complexity.
The first relatively practical method is cylindrical algebraic decomposition (CAD) algorithm \cite{collins1975quantifier} proposed by Collins in 1975, 
followed by lots of improvements. See for example 
\cite{hong1990improvement,collins1991partial,lazard1994improved,mccallum1998improved,brown2001improved}.
Unfortunately, those variants do not improve the complexity of the original algorithm, which is doubly-exponential.
On the other hand, SMT(NRA) is important in theorem proving and program verification, since most complicated programs use real variables and perform nonlinear arithmetic operation on them.
Particularly, SMT(NRA) has various applications in the formal analysis of hybrid systems, dynamical systems and probabilistic systems (see the book \cite{clarke2018handbook} for reference).

The most popular approach for solving SMT(NRA) is the lazy approach, also known as DPLL(T) \cite{biere2009handbook}.
It combines a propositional satisfiability (SAT) solver that uses a conflict-driven clause
learning (CDCL) style algorithm to find assignments of the propositional abstraction of a polynomial formula and a theory solver that checks the consistency of sets of polynomial constraints.
The effort in the approach is devoted to both aspects.
For the theory solver, the only complete method is the CAD method, and 
there also exist many efficient but incomplete methods, such as linearisation \cite{cimatti2018incremental}, interval constraint propagation \cite{tung2016rasat} and virtual substitution \cite{weispfenning1997quantifier}.
Recall that the complexity of the CAD method is doubly-exponential, which makes the cost of simply using it as a theory solver unacceptable.
In order to ease the burden of using CAD, 
an improved CDCL-style search framework, the model constructing satisfiability calculus (MCSAT) framework \cite{jovanovic2012solving,de2013model}, was proposed.
Further, 
there are many optimizations on CAD projection operation, {\it e.g.} 
\cite{brown2015constructing,haokun2020arXiv,abraham2021deciding,DBLP:journals/corr/abs-2212-09309}, custom-made for this framework.

The development of this approach brings us effective SMT(NRA) solvers.
Almost all state-of-the-art SMT(NRA) solvers are based on the lazy approach, including Z3 \cite{moura2008z3}, CVC5 \cite{barbosa2022cvc5}, Yices2 \cite{dutertre2014yices} and MathSAT5 \cite{cimatti2013mathsat5}. 
These solvers have made great process to SMT(NRA).
However, the
time and memory usage of them on some hard instances may be unacceptable,
particularly when the proportion of nonlinear polynomials in all polynomials appearing in the formula is high.
It pushes us to design algorithms which perform well on these hard instances. 

Local search is an incomplete method for solving optimization problems.
A local search algorithm moves from solution to solution in the space of candidate solutions (the search space) by applying local changes, until an optimal solution is found or a time bound is reached.
It is well known that local search method has been successfully applied to SAT problems.
Recent years, some efforts trying to develop local search method for SMT solving are inspiring:
Under the DPLL(T) framework, 
Griggio et al. \cite{griggio2011stochastic} introduced a general procedure for integrating a local search solver of the WalkSAT family with a theory solver.
Pure local search algorithms \cite{frohlich2015stochastic,niemetz2016precise} were proposed to solve SMT problems with respect to the theory of bit-vectors directly on the theory level. 
Cai et al. \cite{cai2022local} developed a local search procedure for SMT on the theory of linear integer arithmetic (LIA) through the \emph{critical move} operation, which
works on the literal-level and changes the value of one variable in a false LIA literal to make it true. 
We also notice that there exists a local search SMT solver for the theory of NRA, called NRA-LS, performing well at the SMT Competition 2022\footnote{\url{https://smt-comp.github.io/2022}.}.
A simple description of the solver without details about local search can be found in \cite{NRALS2022}.

In this paper, we propose a local search algorithm for a special subclass of SMT(NRA), 
where all constraints are strict inequalities. 
The idea of applying the local search method to SMT(NRA) comes from CAD, which is a decomposition of the search space $\R^{n}$ into finitely many cells such that every polynomial in the formula is sign-invariant on each cell.
CAD guarantees that the search space only has finitely many states.
Thus, the search space for SMT(NRA) is completely similar to that for SAT
so that we may use a local search framework for SAT to solve SMT(NRA).

The key point is to define an operation, like \emph{flip} for SAT, to perform local changes. 
We propose a novel operation, called \emph{cell-jump}, 
updating the current assignment $x_1\mapsto a_1,\ldots,x_n\mapsto a_n~(a_i\in\Q)$ to a solution of a false polynomial constraint `$p<0$' or `$p>0$'.
Different from the critical move operation for linear integer constraints,
it is difficult to determine the threshold value of some variable $x_i$ such that the false polynomial constraint becomes true.
We deal with the issue by the method of real root isolation, which fixes every real root of the univariate polynomial $p(a_1,\ldots,a_{i-1},x_i,a_{i+1},\ldots,a_n)$ in an open interval sufficiently small with rational endpoints.
If there exists at least one 
endpoint making the false constraint true, a cell-jump operation assigns $x_i$ to 
one closest to $a_i$.
The procedure can be viewed as searching for a solution 
along a line parallel to the $x_i$-axis.
We further define a cell-jump operation that searches along any fixed straight line, and thus one cell-jump may change the values of more than one variables.
Each step, the local search algorithm picks a cell-jump operation to execute according to a two-level operation selection and updates the current assignment, until a solution to the polynomial formula is found or 
the terminal condition is satisfied.
Moreover, our algorithm can be generalized to deal with a wider subclass of SMT(NRA) where polynomial equations linear w.r.t. some variable are allowed.

The local search algorithm is implemented with {\tt Maple2022} as a tool.
Experiments are conducted to evaluate the tool
on two classes of benchmarks, including selected instances 
from SMTLIB\footnote{\url{https://smtlib.cs.uiowa.edu/benchmarks.shtml}.},
and some hard instances generated randomly with only nonlinear constraints.
Experimental results show that our tool is competitive with state-of-the-art SMT solvers on the SMTLIB benchmarks, and performs particularly well
on the hard instances. 
We also combine our tool with CVC5 to obtain
a sequential portfolio solver, which shows better performance.


The rest of the paper is organized as follows.
The next section introduces some basic definitions and notation and 
a general local search framework for solving a satisfiability problem.
Section \ref{sec:finite_states} shows from the CAD perspective, the search space for SMT(NRA) only has finite states.
In Section \ref{sec:cell_jump}, we describe the cell-jump operation, while in Section \ref{sec:score_func} we provide the scoring function. 
The main algorithm is presented in Section \ref{sec:main_alg}. 
And in Section \ref{sec:experiment}, experimental results are provided to indicate the efficiency of the algorithm. 
Finally, the paper is concluded in Section \ref{sec:conclusion}.
\vspace{-2mm}
\section{Preliminaries}\label{sec:pre}
\vspace{-2mm}
\subsection{Notation}
\vspace{-1mm}
Let $\vX:=(x_1,\ldots,x_n)$ be a vector of variables.
Denote by $\Q$, $\R$ and $\Z$ the set of rational numbers, real numbers and integer numbers, respectively.
Let $\Q[\vX]$ and $\R[\vX]$ be the ring of polynomials in the variables $x_1,\ldots,x_n$ with coefficients in $\Q$ and in $\R$, respectively.

\vspace{-2mm}
\begin{definition}[\bf Polynomial Formula]
Suppose $\Lambda=\{P_1,\ldots,P_m\}$ where every $P_i$ is a nonempty finite subset of $\Q[\vX]$. 
The following formula
\begin{align*}
    F=\bigwedge_{P_i\in\Lambda}\bigvee_{p_{ij}\in P_i}p_{ij}(x_1,\ldots,x_n)\rhd_{ij}0,~{\rm where}~\rhd_{ij}\in\{<,>,=\},
\end{align*}
is called a \emph{polynomial formula}.
Additionally, we call
$p_{ij}(x_1,\ldots,x_n)\rhd_{ij}0$ an \emph{atomic polynomial formula}, and $\bigvee_{p_{ij}\in P_j}p_{ij}(x_1,\allowbreak\ldots,x_n)\rhd_{ij}0$ a \emph{polynomial clause}.
\end{definition}
\vspace{-2mm}


For any polynomial formula $F$, $\poly(F)$ denotes the set of polynomials appearing in $F$.
For any atomic formula $\ell$, 
$\poly(\ell)$ denotes the polynomial appearing in $\ell$ and $\rela(\ell)$ denotes the relational operator (`$<$', `$>$' or `$=$') of $\ell$.

For any polynomial formula $F$, an \emph{assignment} is a mapping $\alpha:\vX\rightarrow\R^n$ such that $\alpha(\vX)=(a_1,\ldots,a_n)$ where $a_i\in\R$.
Given an assignment $\alpha$,
\vspace{-1mm}
\begin{itemize}
    \item an atomic polynomial formula is \emph{true under $\alpha$} if it evaluates to true under $\alpha$, and otherwise it is \emph{false under $\alpha$},
    \item a polynomial clause is \emph{satisfied under $\alpha$} if
    at least one atomic formula in the clause is true under $\alpha$, and \emph{falsified under $\alpha$} otherwise. 
\end{itemize}
\vspace{-1mm}
When the context is clear, we simply say a \emph{true} (or \emph{false}) atomic polynomial formula and a \emph{satisfied} (or \emph{falsified}) polynomial clause.
A polynomial formula is \emph{satisfied} if there exists an assignment $\alpha$ such that all clauses in the formula are satisfied under $\alpha$, and such an assignment is a \emph{solution} to the polynomial formula.
A polynomial formula is \emph{unsatisfied} if any assignment is not a solution.

\vspace{-4mm}
\subsection{A General Local Search Framework}\label{sec:general_LS}
\vspace{-1mm}


When applying local search algorithms to solve a satisfiability problem, the search space is the set of all assignments.
A general local search framework begins with an assignment.
Every time, one of the operations with the greatest score is picked and the assignment is updated after executing the operation until reaching the set terminal condition. 
Below, we give the formal definitions of \emph{operation} and \emph{scoring function}.
\vspace{-1mm}

\begin{definition}[\bf Operation]
Let $F$ be a formula and $\alpha$ be an assignment which is not a solution of $F$.
An \emph{operation} modifies $\alpha$ to another assignment $\alpha'$.
\end{definition}
\vspace{-4mm}




\begin{definition}[\bf Scoring Function]\label{def:score_func}
Let $F$ be a formula.
Suppose $\alpha$ is the current assignment and
$op$ is an operation. 
A \emph{scoring function} is defined as $\score(op,\alpha):=\cost(\alpha)-\cost(\alpha')$,
where the real-valued function $\cost$ measures the cost of making $F$ satisfied under an assignment according to some heuristic, and $\alpha'$ is the assignment after executing $op$.
\end{definition}
\vspace{-4mm}

\begin{example}
In local search algorithms for SAT, a standard operation is \emph{flip},
which modifies the current assignment by flipping the value of one Boolean variable from false to true or vice-versa. 
A commonly used scoring function 
measures the change on the number of falsified clauses by flipping a variable.
Thus, operation $op$ is $\flip(b)$ for some Boolean variable $b$, 
and $\cost(\alpha)$ is interpreted as the number of falsified clauses under the assignment $\alpha$.
\end{example}
\vspace{-2mm}

Actually, only when $\score(op,\alpha)$ is a positive number, does it make sense to execute operation $op$, since the operation guides the current assignment to an assignment with less cost of being a solution.
\vspace{-2mm}

\begin{definition}[\bf Decreasing Operation]\label{def:de_op}
Suppose $\alpha$ is the current assignment.
Given a scoring function $\score$,
an operation $op$ is a \emph{decreasing operation} under $\alpha$ if $\score(op,\alpha)>0$.
\end{definition}
\vspace{-2mm}

A general local search framework is described in Algorithm \ref{alg:LocalSearchProc}.
The framework was used in GSAT  \cite{mitchell1992new} for solving SAT problems.
Remark that if the input formula $F$ is unsatisfied, Algorithm \ref{alg:LocalSearchProc} outputs ``unknown''.
If it is satisfied, Algorithm \ref{alg:LocalSearchProc} outputs either (i) a solution of $F$ if the solution is found successfully, or (ii) ``unknown'' if the algorithm fails.

\vspace{-7mm}
\begin{algorithm}[h]
\scriptsize
\DontPrintSemicolon
\LinesNumbered
\SetKwInOut{Input}{Input}
\SetKwInOut{Output}{Output}
\Input{a formula $F$ and a terminal condition $\varphi$
}
\Output{a solution to $F$ or unknown
}
\caption{{\bf General Local Search Framework}}\label{alg:LocalSearchProc}
\BlankLine
initialize assignment $\alpha$


\While{the terminal condition $\varphi$ is not satisfied}
{
\eIf{$\alpha$ satisfies $F$}
{\Return{$\alpha$}}
{$op\leftarrow$ one of the decreasing operations with the greatest score

perform $op$ to modify $\alpha$
}





}
\Return{unknown}
\end{algorithm}
\vspace{-7mm}

\vspace{-4mm}
\section{The Search Space of SMT(NRA)}\label{sec:finite_states}
\vspace{-2mm}
%
%
The search space for SAT problems consists of finitely many assignments. So, theoretically speaking, a local search algorithm can eventually find a solution, as long as the formula indeed has a solution and there is no cycling during the search. 
It seems intuitively, however, that the search space of an SMT(NRA) problem, {\it e.g.} $\R^{n}$, is infinite and thus search algorithms may not work. 

Fortunately, due to Tarski's work and the theory of CAD, SMT(NRA) is decidable. 
Given a polynomial formula in $n$ variables, by the theory of CAD, $\R^{n}$ is decomposed into finitely many cells such that every polynomial in the formula is sign-invariant on each cell. Therefore, the search space of the problem is essentially finite. The cells of SMT(NRA) are very similar to the Boolean assignments of SAT, so just like traversing all Boolean assignments in SAT, there exists a basic strategy to traverse all cells.

In this section, we describe the search space of SMT(NRA) based on the CAD theory from a local search perspective, providing a theoretical foundation for the operators and heuristic frameworks we will propose in the next sections.

\vspace{-1mm}

\begin{example}\label{ex:twoellipses} 
Consider the polynomial formula
\[F\;=\;(f_1 > 0  \lor  f_2 > 0) \land (f_1 < 0 \lor f_2 < 0),\]
where $f_1 = 17x^2 + 2xy + 17y^2 + 48x - 48y$ and $f_2=17x^2 - 2xy + 17y^2 - 48x - 48y.$

The solution set of $F$ is shown as the shaded area in Figure \ref{fig:twoellipses}. 
Notice that $\poly(F)$
consists of two polynomials and decomposes $\R^2$ into $10$ areas: $C_1,\ldots,C_{10}$ (see Figure \ref{fig:twoellipses2}). 
We refer to these areas as \emph{cells}.

\vspace{-6mm}
\begin{figure}[h]
\begin{minipage}[htbp]{0.45\linewidth}
\begin{subfigure}
    \centering
    \begin{tikzpicture}[scale=0.5,line cap=round,line join=round,>=triangle 45,x=1cm,y=1cm]
\begin{axis}[
x=1cm,y=1cm,
axis lines=middle,
xmin=-4.5,
xmax=4.5,
ymin=-2.5,
ymax=4.5,
xtick={-5,-4,...,5},
ytick={-2,-1,...,4},]
\clip(-4.5,-2.5) rectangle (4.5,4.5);
\draw (1.4062674680691189,5.97677986476334) node[scale=3] {$f_2$} ;
\def\ellia{(-1.5,1.5) circle[x radius=2.121320343559643, y radius=2,rotate=-45]}
\def\ellib{(1.5,1.5) circle[x radius=2.121320343559643, y radius=2,rotate=45]}

\draw [line width=2pt] \ellia;
\draw [line width=2pt] \ellib;
\fill[pattern=north west lines ,even odd rule] \ellia \ellib;

\draw (4.3,0.2)  node[scale=1] {$x$};
\draw (-0.2,4.3)  node[scale=1] {$y$};

\end{axis}
\end{tikzpicture}
    \vspace{-5mm}
    \caption{The solution set of $F$ in Example \ref{ex:twoellipses}.
    }
    \label{fig:twoellipses}
\end{subfigure}
\end{minipage}
\hfill
\begin{minipage}[htbp]{0.45\linewidth}
\begin{subfigure}
    \centering
    \begin{tikzpicture}[scale=0.5,line cap=round,line join=round,>=triangle 45,x=1cm,y=1cm]
\begin{axis}[
x=1cm,y=1cm,
axis lines=middle,
xmin=-4.5,
xmax=4.5,
ymin=-2.5,
ymax=4.5,
xtick={-5,-4,...,5},
ytick={-2,-1,...,4},]
\clip(-4.5,-2.5) rectangle (4.5,4.5);
\draw (1.4062674680691189,5.97677986476334) node[scale=3] {$f_2$} ;
\def\ellia{(-1.5,1.5) circle[x radius=2.121320343559643, y radius=2,rotate=-45]}
\def\ellib{(1.5,1.5) circle[x radius=2.121320343559643, y radius=2,rotate=45]}
\fill (0,0) circle (2.5pt);
\fill (0,2.82) circle (2.5pt);
\draw (2,1)  node[scale=1] {$C_1$};
\draw (-2,1)  node[scale=1] {$C_2$};
\draw (0.2,1.2)  node[scale=1] {$C_3$};

\draw (2,4)  node[scale=1] {$C_4$};
\draw (-3,2) node[scale=1] {$C_5$};
\draw [-latex,line width=0.5pt]  (-3.1,2.1) -- (-3.3,2.3);
\draw (3,2) node[scale=1] {$C_6$};
\draw [-latex,line width=0.5pt]  (3.1,2.1) -- (3.3,2.3);
\draw (-1, 0.6) node[scale=1] {$C_7$};
\draw [-latex,line width=0.5pt]  (-0.8,0.6) -- (-0.4,0.6);
\draw (1, 0.6) node[scale=1] {$C_8$};
\draw [-latex,line width=0.5pt]  (0.8,0.6) -- (0.4,0.6);
\draw (0.4,-0.7) node[scale=1] {$C_9$};
\draw [-latex,line width=0.5pt]  (0.3,-0.6) -- (0.05,-0.1);
\draw (0.7,2.82) node[scale=1] {$C_{10}$};
\draw [-latex,line width=0.5pt]  (0.4,2.82) -- (0.1,2.82);

\draw [line width=2pt] \ellia;
\draw [line width=2pt] \ellib;

\draw (4.3,0.2)  node[scale=1] {$x$};
\draw (-0.2,4.3)  node[scale=1] {$y$};

\end{axis}
\end{tikzpicture}
    \vspace{-5mm}
    \caption{The zero level set of $\poly(F)$ decomposes $\R^2$ into $10$ cells.}
    \label{fig:twoellipses2}
\end{subfigure}
\end{minipage}

\end{figure}
\end{example}
\vspace{-11mm}

\begin{definition}[Cell]\label{def:cell}
For any finite set $Q\subseteq\R[\vX]$, a \emph{cell} of $Q$ is
a maximally connected set in $\R^{n}$ on which the sign of every polynomial in $Q$ is constant.
For any point $\bar{a}\in\R^{n}$,
we denote by $\cell{Q}{\bar{a}}$ the cell of $Q$ containing $\bar{a}$. 
\end{definition}
\vspace{-2mm}

By the theory of CAD, we have 
\vspace{-2mm}
\begin{corollary}
For any finite set $Q\subseteq \R[\vX]$, the number of cells of $Q$ is finite.
\end{corollary}
\vspace{-2mm}

It is obvious that any two cells of $Q$ are disjoint and the union of all cells of $Q$ equals $\R^{n}$.
Definition \ref{def:cell} shows that
for a polynomial formula $F$ with $\poly(F)=Q$,
the satisfiability of $F$ is constant on every cell of $Q$, that is, either all the points in a cell are  solutions to $F$ or none of them are solutions to $F$.

\vspace{-2mm}

\begin{example}\label{ex:twoellipses2} 

Consider the polynomial formula $F$
in Example \ref{ex:twoellipses}. 
As shown in Figure \ref{fig:twoellipses3}, assume that we start from point $a$ to search for a solution to $F$.
Jumping from $a$ to $b$ makes no difference, as both points are in the same cell and thus neither are solutions to $F$.
However, jumping from $a$ to $c$ or from $a$ to $d$ 
crosses different cells and we may discover a cell satisfying $F$.
Herein, the cell containing $d$ satisfies $F$.
\begin{figure}[h]
    \begin{minipage}[htbp]{0.45\linewidth}
        \begin{subfigure}
            \centering
            \begin{tikzpicture}[scale=0.5,line cap=round,line join=round,>=triangle 45,x=1cm,y=1cm]
\begin{axis}[
x=1cm,y=1cm,
axis lines=middle,
xmin=-4.5,
xmax=4.5,
ymin=-2.5,
ymax=4.5,
xtick={-5,-4,...,5},
ytick={-2,-1,...,4},]
\clip(-4.5,-2.5) rectangle (4.5,4.5);
\draw (1.4062674680691189,5.97677986476334) node[scale=3] {$f_2$} ;
\def\ellia{(-1.5,1.5) circle[x radius=2.121320343559643, y radius=2,rotate=-45]}
\def\ellib{(1.5,1.5) circle[x radius=2.121320343559643, y radius=2,rotate=45]}

\draw [line width=2pt] \ellia;
\draw [line width=2pt] \ellib;

\draw (2,4)   circle (2.5pt);
\draw (2.2,4.2)  node[scale=1] {$a$};
\draw (0,4)   circle (2.5pt);
\draw (0.2,4.2)  node[scale=1] {$b$};
\draw [-latex,line width=0.5pt]  (1.8,4) -- (1.2,4);
\draw (2,3.53)   circle (2.5pt);
\draw (2.2,3.7)  node[scale=1] {$c$};
\draw [-latex,line width=0.5pt]  (2,3.9) -- (2,3.6);
\draw (2.2,0.2)  node[scale=1] {$d$};

\draw [-latex,line width=0.5pt]  (2,3.4) -- (2,0.3);
\fill (0,0) circle (2.5pt);
\draw (0,-2)   circle (2.5pt);
\draw (0,2)   circle (2.5pt);
\draw (0,2.82)   circle (2.5pt);

\fill (2,0) circle (2.5pt);
\draw (2,-0.47)   circle (2.5pt);
\draw (2,-2)   circle (2.5pt);

\fill (2.82,0) circle (2.5pt);
\draw (2.82,-2)   circle (2.5pt);
\draw (2.82,2)   circle (2.5pt);
\draw (2.82,3.16)   circle (2.5pt);
\draw (2.82,4)   circle (2.5pt);

\fill (4,0) circle (2.5pt);
\fill (-2,0) circle (2.5pt);
\draw (-2,-0.47)   circle (2.5pt);
\draw (-2,-2)   circle (2.5pt);
\draw (-2,4)   circle (2.5pt);
\draw (-2,3.53)   circle (2.5pt);

\fill (-2.82,0) circle (2.5pt);
\draw (-2.82,-2)   circle (2.5pt);
\draw (-2.82,2)   circle (2.5pt);
\draw (-2.82,3.16)   circle (2.5pt);
\draw (-2.82,4)   circle (2.5pt);

\fill (-4,0) circle (2.5pt);

\draw (4.3,0.2)  node[scale=1] {$x$};
\draw (-0.2,4.3)  node[scale=1] {$y$};

\end{axis}
\end{tikzpicture}
            \vspace{-5mm}
            \caption{Jumping from point $a$ to search for a solution of $F$.
            }
            \label{fig:twoellipses3}
        \end{subfigure}
    \end{minipage}
    \hfill
    \begin{minipage}[htbp]{0.45\linewidth}
        \begin{subfigure}
            \centering
            \begin{tikzpicture}[scale=0.5,line cap=round,line join=round,>=triangle 45,x=1cm,y=1cm]
\begin{axis}[
x=1cm,y=1cm,
axis lines=middle,
xmin=-4.5,
xmax=4.5,
ymin=-2.5,
ymax=4.5,
xtick={-5,-4,...,5},
ytick={-2,-1,...,4},]
\clip(-4.5,-2.5) rectangle (4.5,4.5);
\draw (1.4062674680691189,5.97677986476334) node[scale=3] {$f_2$} ;
\def\ellia{(-1.5,1.5) circle[x radius=2.121320343559643, y radius=2,rotate=-45]}
\def\ellib{(1.5,1.5) circle[x radius=2.121320343559643, y radius=2,rotate=45]}

\draw [line width=2pt] \ellia;
\draw [line width=2pt] \ellib;

   
\draw [line width=2pt]  (-3.56,-3) -- (-3.56,6);
\draw [line width=2pt]  (-0.56,-3) -- (-0.56,6);
\draw [line width=2pt]  (0,-3) -- (0,6);
\draw [line width=2pt]  (0.56,-3) -- (0.56,6);
\draw [line width=2pt]  (3.56,-3) -- (3.56,6);

\draw (4.3,0.2)  node[scale=1] {$x$};
\draw (-0.2,4.3)  node[scale=1] {$y$};

\draw (2,4)   circle (2.5pt);
\draw (0,4)   circle (2.5pt);
\draw (2,3.53)   circle (2.5pt);

\draw (0,0) circle (2.5pt);
\draw (0,-1)   circle (2.5pt);
\draw (0,2)   circle (2.5pt);
\draw (0,2.82)   circle (2.5pt);

\draw (2,0) circle (2.5pt);
\draw (2,-0.47)   circle (2.5pt);
\draw (2,-1)   circle (2.5pt);

\draw (2.82,0) circle (2.5pt);
\draw (2.82,-1)   circle (2.5pt);
\draw (2.82,1.5)   circle (2.5pt);
\draw (2.82,3.16)   circle (2.5pt);
\draw (2.82,4)   circle (2.5pt);

\draw (4,0) circle (2.5pt);
\draw (-4,0) circle (2.5pt);

\draw (-2,0) circle (2.5pt);
\draw (-2,-0.47)   circle (2.5pt);
\draw (-2,-1)   circle (2.5pt);
\draw (-2,4)   circle (2.5pt);
\draw (-2,3.53)   circle (2.5pt);

\draw (-2.82,0) circle (2.5pt);
\draw (-2.82,-1)   circle (2.5pt);
\draw (-2.82,1.5)   circle (2.5pt);
\draw (-2.82,3.16)   circle (2.5pt);
\draw (-2.82,4)   circle (2.5pt);

\draw (-0.56,-1)   circle (2.5pt);
\draw (-0.56,-0.39)   circle (2.5pt);
\draw (-0.56,0) circle (2.5pt);
\draw (-0.56,1.31)   circle (2.5pt);
\draw (-0.56,2.5)   circle (2.5pt);
\draw (-0.56,3.28)   circle (2.5pt);
\draw (-0.56,4)   circle (2.5pt);

\draw (0.56,-1)   circle (2.5pt);
\draw (0.56,-0.39)   circle (2.5pt);
\draw (0.56,0) circle (2.5pt);
\draw (0.56,1.31)   circle (2.5pt);
\draw (0.56,2.5)   circle (2.5pt);
\draw (0.56,3.28)   circle (2.5pt);
\draw (0.56,4)   circle (2.5pt);

\draw (-3.56,0)   circle (2.5pt);
\draw (-3.56,1.59)   circle (2.5pt);
\draw (-3.56,4)   circle (2.5pt);
\draw (3.56,0)   circle (2.5pt);
\draw (3.56,1.59)   circle (2.5pt);
\draw (3.56,4)   circle (2.5pt);

\draw (-3.2,0)   circle (2.5pt);
\draw (-3.2,0.44)   circle (2.5pt);
\draw (-3.2,1.5)   circle (2.5pt);
\draw (-3.2,2.73)   circle (2.5pt);
\draw (-3.2,4)   circle (2.5pt);
\draw (3.2,0)   circle (2.5pt);
\draw (3.2,0.44)   circle (2.5pt);
\draw (3.2,1.5)   circle (2.5pt);
\draw (3.2,2.73)   circle (2.5pt);
\draw (3.2,4)   circle (2.5pt);

\draw (-0.3,-1)   circle (2.5pt);
\draw (-0.3,-0.25)   circle (2.5pt);
\draw (-0.3,0)   circle (2.5pt);
\draw (-0.3,0.39)   circle (2.5pt);
\draw (-0.3,1.3)   circle (2.5pt);
\draw (-0.3,2.39)   circle (2.5pt);
\draw (-0.3,2.7)   circle (2.5pt);
\draw (-0.3,3.1)   circle (2.5pt);
\draw (-0.3,4)   circle (2.5pt);
\draw (0.3,-1)   circle (2.5pt);
\draw (0.3,-0.25)   circle (2.5pt);
\draw (0.3,0)   circle (2.5pt);
\draw (0.3,0.39)   circle (2.5pt);
\draw (0.3,1.3)   circle (2.5pt);
\draw (0.3,2.39)   circle (2.5pt);
\draw (0.3,2.7)   circle (2.5pt);
\draw (0.3,3.1)   circle (2.5pt);
\draw (0.3,4)   circle (2.5pt);

\end{axis}
\end{tikzpicture}
            \vspace{-5mm}
            \caption{A cylindrical expansion of a cylindrically complete set containing $\poly(F)$.}
            \label{fig:twoellipses4}
        \end{subfigure}
    \end{minipage}
\end{figure}
\end{example}

For the remainder of this section, we will demonstrate how to traverse all cells through point jumps between cells. 
The method essentially traversing cell by cell in a variable by variable direction will be explained step by step from Definition \ref{def:expansion} to Definition \ref{def:cyl_complete}. 
\begin{definition}[Expansion]\label{def:expansion}
Let $Q\subseteq\R[\vX]$ be finite and $\bar{a}=(a_1,\ldots,a_n)\in \R^n$. 
Given a variable $x_i~(1\le i\le n)$, let $r_1<\cdots<r_s$ be all real roots of $\{q(a_1,\ldots,a_{i-1},x_i,\allowbreak a_{i+1},\ldots,a_n)\mid q(a_1,\ldots,a_{i-1},x_i,\allowbreak a_{i+1},\ldots,a_n)\not\equiv 0,\  q\in Q\}$,
where $s\in\Z_{\ge0}$. 
An \emph{expansion} of $\bar{a}$ to $x_i$ on $Q$ is a point set $\Lambda\subseteq \R^n$ satisfying 

\vspace{-2mm}
\begin{enumerate}[label=$($\alph*$)$]
    \item $\bar{a}\in \Lambda$ and  $(a_1,\ldots,a_{i-1},r_j,a_{i+1},\ldots,a_n)\in \Lambda$ for $1\leq j\leq s$,
    \item for any $\bar{b}=(b_1,...,b_n)\in \Lambda$,
    $b_j=a_j$ for $j\in\{1,\ldots,n\}\setminus\{i\}$, and
    \item
    for any interval  $I\in\{(-\infty,r_1),(r_1,r_2),\ldots,(r_{s-1},r_s),(r_s,+\infty)\}$,
    there exists unique $\bar{b}=(b_1,...,b_n)\in \Lambda$ such that $b_i\in I.$ 
\end{enumerate}
\vspace{-2mm}
For any point set $\{\vA^{(1)},\ldots,\vA^{(m)}\}\subseteq\R^{n}$, an \emph{expansion} of the set to $x_i$ on $Q$ is $\bigcup_{j=1}^m\Lambda_j$, where $\Lambda_j$ is an expansion of $\vA^{(j)}$ to $x_i$ on $Q$.
\end{definition} 
\vspace{-3mm}
\begin{example}\label{ex:expansion} 
Consider the polynomial formula $F$
in Example \ref{ex:twoellipses}.  
The set of black solid points in Figure \ref{fig:twoellipses3}, denoted as $\Lambda$, is an expansion of point $(0,0)$ to $x$ on $\poly(F)$.
The set of all points (including black solid points and hollow points) is an expansion of $\Lambda$ to $y$ on $\poly(F)$.
\end{example}
\vspace{-2mm}

As shown in Figure \ref{fig:twoellipses3},  an expansion of a point   to some variable  is actually a result of the point continuously jumping to adjacent cells along that variable direction.
Next, we describe the expansion of all variables in order, which is a result of jumping from cell to cell along the directions of variables w.r.t. a variable order.
\vspace{-2mm}
\begin{definition}[Cylindrical Expansion]\label{def:cyl_expansion}
Let $Q\subseteq\R[\vX]$ be finite and $\bar{a}\in \R^n$. 
Given a variable order $x_1\prec\cdots\prec x_n$, 
a \emph{cylindrical expansion} of $\bar{a}$ w.r.t. the variable order on $Q$
is $\bigcup_{i=1}^n \Lambda_i$, where 
$\Lambda_1$ is an expansion of $\bar{a}$ to $x_1$ on $Q$, and
for $1\leq i<n$, $\Lambda_{i+1}$ is an expansion of $\Lambda_{i}$ to $x_{i+1}$ on $Q$.
When the context is clear, we simply call
$\bigcup_{i=1}^n \Lambda_i$ a \emph{cylindrical expansion} of $Q$.
\end{definition}
\vspace{-3mm}
\begin{example}\label{ex:cyl_expansion} 
Consider the formula $F$
in Example \ref{ex:twoellipses}.  
It is clear that the set of all points 
in Figure \ref{fig:twoellipses3} is a cylindrical expansion of point $(0,0)$ w.r.t. $x\prec y$ on $\poly(F)$.
The expansion actually describes the following jumping process. First, the origin $(0,0)$ jumps along the $x$-axis to the black points, and then those black points jump along the $y$-axis direction to the white points.
\end{example}
\vspace{-2mm}

Clearly, a cylindrical expansion is 
similar to how a Boolean vector is flipped variable by variable. Note that the points in the expansion in Figure \ref{fig:twoellipses3} do not cover all the cells ({\it e.g.} $C_7$ and $C_8$ in Figure \ref{fig:twoellipses2}), but if we start from $(0,2)$, all the cells can be covered. This implies that whether all the cells can be covered depends on the starting point. 

\vspace{-2mm}
\begin{definition}[Cylindrically Complete]\label{def:cyl_complete}
Let $Q\subseteq\R[\vX]$ be finite.
Given a variable order $x_1\prec\cdots\prec x_n$, 
$Q$ is said to be \emph{cylindrically complete} w.r.t. the variable order,
if for any $\vA\in \R^n$ and for any cylindrical expansion $\Lambda$ of $\vA$ w.r.t. the order on $Q$,
every cell of $Q$ contains at least one point in $\Lambda$. 
\end{definition}
\vspace{-4mm}
\begin{theorem}\label{thm:CAD}
For any finite set $Q\subseteq\R[\vX]$ and any variable order, 
there exists 
$Q'$ such that $Q\subseteq Q'\subseteq\R[\vX]$ and $Q'$ is cylindrically complete w.r.t. the variable order.
\end{theorem}
\vspace{-4mm}

\begin{proof} 
Let $Q'$ be the projection set of $Q$ \cite{collins1975quantifier,mccallum1998improved,brown2001improved} obtained from the CAD projection operator w.r.t. the variable order.
According to the theory of CAD, $Q'$ is cylindrically complete.\hfill$\square$
\end{proof}
\vspace{-4mm}

\begin{corollary}\label{coro:CAD}
For any polynomial formula $F$ and any variable order,
there exists a finite set $Q\subseteq\R[\vX]$ such that  for any cylindrical expansion $\Lambda$ of $Q$, every cell of $\poly(F)$ contains at least one point in $\Lambda$. 
Furthermore, $F$ is satisfiable if and only if 
$F$ has solutions in  $\Lambda$.
\end{corollary}
\vspace{-4mm}

\begin{example} 
Consider the polynomial formula $F$
in Example \ref{ex:twoellipses}.
By the proof of Theorem \ref{thm:CAD}, $Q':=\{
x, -2 - 3 x + x^2,\; -2 + 3 x + x^2,\; 
 10944 + 17 x^2,\; 
f_1,\; f_2\}$ is a cylindrically complete set w.r.t. $x\prec y$
containing $\poly(F)$.
As shown in Figure \ref{fig:twoellipses4}, the set of all (hollow) points is a cylindrical expansion of point $(0,0)$ w.r.t. $x \prec y$ on $Q'$, which 
covers all cells of $\poly(F)$.
\end{example}
\vspace{-2mm}

Corollary \ref{coro:CAD} shows that for a polynomial formula $F$, there exists a finite set $Q\subseteq\R[\vX]$ such that we can traverse all the cells of $\poly(F)$ through a search path containing all points in a cylindrical expansion of $Q$.  The cost of traversing the cells is very high, and in the worst case, the number of cells will grow exponentially with the number of variables.

The key to building a local search on SMT(NRA) is to construct a heuristic search based on the operation of jumping between cells. 

\vspace{-4mm}
\section{The Cell-Jump Operation}\label{sec:cell_jump}
\vspace{-2mm}

In this section, we propose a novel operation, called \emph{cell-jump}, 
that performs local changes in our algorithm.
The operation is determined by the means of real root isolation.
We review the method of real root isolation and define \emph{sample points} in Section \ref{subsec:realrootiso}.
Section \ref{subsec:celljump} and Section \ref{subsec:cell_jump_dir} present a cell-jump operation along a line parallel to a coordinate axis and along any fixed straight line, respectively.
\vspace{-5mm}
\subsection{Sample Points}\label{subsec:realrootiso}
\vspace{-2mm}
Real root isolation is a symbolic way to compute the real roots of a polynomial, which is of fundamental importance in computational real algebraic geometry ({\it e.g.}, it is a routing sub-algorithm for CAD).
There are many efficient algorithms and popular tools in computer algebra systems such as  Maple and Mathematica to isolate the real roots of polynomials.

We first introduce the definition of \emph{sequences of isolating intervals} for nonzero univariate polynomials,
which can be obtained by any real root isolation tool. 
\begin{definition}[\bf Sequence of Isolating Intervals]
For any nonzero univariate polynomial $p(x)\in\Q[x]$, a \emph{sequence of isolating intervals} of $p(x)$ is a sequence of open intervals $(a_1,b_1),\ldots,(a_s,b_s)$ where $s\in\Z_{\ge0}$, such that
\vspace{-1.5mm}
\begin{enumerate}
    \item for each $i~(1\le i\le s)$, $a_i,b_i\in\Q$, $a_i< b_i$ and $b_i < a_{i+1}$, 
    \item each interval $(a_i,b_i)~(1\le i\le s)$ has exactly one real root of $p(x)$, and
    \item none of the real roots of $p(x)$ are in $\R\setminus\bigcup_{i=1}^{s}(a_i,b_i)$.
\end{enumerate}
\vspace{-1.5mm}
Specially, the sequence of isolating intervals is empty, {\it i.e.}, $s=0$, when $p(x)$ has no real roots.
\end{definition}
\vspace{-2mm}

By means of sequences of isolating intervals, we define \emph{sample points} of univariate polynomials, which is the key concept of the \emph{cell-jump} operation proposed in Section \ref{subsec:celljump} and Section \ref{subsec:cell_jump_dir}. 
\begin{definition}[\bf Sample Point]\label{def:sample_points}
For any nonzero univariate polynomial $p(x)\in\Q[x]$, let $(a_1,b_1),\ldots,(a_s,b_s)$ be a sequence of isolating intervals of $p(x)$ where $s\in\Z_{\ge0}$.
Every point in the set $\{a_1,b_s\}\cup\bigcup_{i=1}^{s-1}\{b_i,\frac{b_i+a_{i+1}}{2},a_{i+1}\}$ is a \emph{sample point} of $p(x)$.
If $x^{*}$ is a sample point of $p(x)$ and $p(x^{*})>0$ $($or $p(x^{*})<0)$, then $x^{*}$ is a \emph{positive sample point} $($or \emph{negative sample point}$)$ of $p(x)$. 
For the zero polynomial, it has no \emph{sample point}, no \emph{positive sample point} and no \emph{negative sample point}.
\end{definition}
\vspace{-4mm}
\begin{remark}\label{remark:about_sample_points}
For any nonzero univariate polynomial $p(x)$ that has real roots,
let $r_1,\ldots,\allowbreak r_s\;(s\in\Z_{\ge1})$ be all distinct real roots of $p(x)$. 
It is obvious that the sign of $p(x)$ is positive constantly or negative constantly on each interval $I$ of the set $\{(-\infty,r_1),(r_1,r_2),\ldots,\allowbreak(r_{s-1},r_s),\allowbreak(r_s,+\infty)\}$.
So, 
we only need to take a point $x^{*}$ from the interval $I$, and then the sign of $p(x^*)$ is the constant sign of $p(x)$ on $I$. 
Specially, we take $a_1$ as the sample point for the interval $(-\infty,r_1)$, $b_i,\frac{b_i+a_{i+1}}{2}~\text{or}~a_{i+1}$ as a sample point for $(r_i,r_{i+1})$ where $1\le i\le s-1$, and
$b_s$ as the sample point for $(r_s,+\infty)$.
By Definition \ref{def:sample_points}, there exists no sample point for the zero polynomial and a univariate polynomial with no real roots.
\end{remark}
\vspace{-4mm}
\begin{example}\label{example:about_sample_points}
Consider the polynomial $p(x)=x^8 - 4x^6 + 6x^4 - 4x^2 + 1$.
It has two real roots $-1$ and $1$, and a sequence of isolating intervals of it is $(-\frac{215}{128}, -\frac{19}{32})$, $(\frac{19}{32}, \frac{215}{128})$.
Every point in the set $\{-\frac{215}{128},-\frac{19}{32},0,\frac{19}{32},\frac{215}{128}\}$ is a sample point of $p(x)$.
Note that $p(x)>0$ holds on the intervals $(-\infty,-1)$ and $(1,+\infty)$, and $p(x)<0$ holds on the interval $(-1,1)$.
Thus, $-\frac{215}{128}$ and $\frac{215}{128}$ are positive sample points of $p(x)$; $-\frac{19}{32},0$ and $\frac{19}{32}$ are negative sample points of $p(x)$.
\end{example}
\vspace{-2mm}

\vspace{-4mm}
\subsection{Cell-Jump Along a Line Parallel to a Coordinate Axis}\label{subsec:celljump}
\vspace{-2mm}
The \emph{critical move} operation \cite[Definition 2]{cai2022local} is a literal-level operation.
For any false LIA literal, the operation changes the value of one variable in it to make the literal true.
In the subsection, we propose a similar operation which adjusts the value of one variable in a false atomic polynomial formula with `$<$' or `$>$'.
\vspace{-1mm}
\begin{definition}
\label{def:cell_jump}
Suppose the current assignment is $\alpha:x_1\mapsto a_1,\ldots, \allowbreak x_n\mapsto a_n$ where $a_i\in\Q$.
Let $\ell$ be a false atomic polynomial formula under $\alpha$ with a relational operator `$<$' or `$>$'.
\vspace{-1mm}
\begin{enumerate}
\item
Suppose $\ell$ is $p(\vX)<0$.
For each variable $x_i$ such that the univariate polynomial $p(a_1,\ldots,a_{i-1},x_i,a_{i+1},\ldots,a_n)$ has negative sample points, there exists a \emph{cell-jump} operation, denoted as $\sm(x_i,\ell)$, assigning $x_i$ to a negative sample point closest to $a_i$.\label{item:def:cell_jump2}
\item
Suppose $\ell$ is $p(\vX)>0$.
For each variable $x_i$ such that the univariate polynomial $p(a_1,\ldots,a_{i-1},x_i,a_{i+1},\ldots,a_n)$ has positive sample points, there exists a \emph{cell-jump} operation, denoted as $\sm(x_i,\ell)$, assigning $x_i$ to a positive sample point closest to $a_i$.\label{item:def:cell_jump1}
\end{enumerate}
\end{definition}
\vspace{-2mm}

Every assignment in the search space can be viewed as a point in $\R^n$.
Then, performing a $\sm(x_i,\ell)$ operation is equivalent to moving one step from the current point $\alpha(\vX)$ along the line $(a_1,\ldots,a_{i-1},\R,a_{i+1},\ldots,a_n)$. 
Since the line is parallel to the $x_i$-axis, we call $\sm(x_i,\ell)$ a \emph{cell-jump along a line parallel to a coordinate axis}.
\vspace{-2mm}
\begin{theorem}\label{thm:cell_jump_xi}
Suppose the current assignment is $\alpha:x_1\mapsto a_1,\ldots, \allowbreak x_n\mapsto a_n$ where $a_i\in\Q$.
Let $\ell$ be a false atomic polynomial formula under $\alpha$ with a relational operator `$<$' or `$>$'.
For every $i~(1\le i\le n)$, 
there exists a solution of $\ell$ in 
$\{\alpha'\mid \alpha'(\vX)\in(a_1,\ldots,a_{i-1},\R,a_{i+1},\ldots,a_n)\}$
if and only if there exists a $\sm(x_i,\ell)$ operation.
\end{theorem}
\vspace{-4mm}
\begin{proof} 
$\Leftarrow$ It is clear by the definition of negative (or positive) sample points.

$\Rightarrow$ Let $S:=\{\alpha'\mid \alpha'(\vX)\in(a_1,\ldots,a_{i-1},\R,a_{i+1},\ldots,a_n)\}$.
It is equivalent to proving that if there exists no $\sm(x_i,\ell)$ operation, then no solution to $\ell$ exists in $S$.
We only prove it for $\ell$ of the form $p(\vX)<0$.
Recall Definition \ref{def:sample_points} and Remark \ref{remark:about_sample_points}.
There are only three cases in which $\sm(x_i,\ell)$ does not exist:
(1) $p^{*}$ is the zero polynomial,
(2) $p^{*}$ has no real roots,
(3) $p^{*}$ has a finite number of real roots, say $r_1,\ldots,r_s~(s\in\Z_{\ge1})$, and $p^{*}$ is positive on $\R\setminus\{r_1,\ldots,r_s\}$, where $p^{*}$ denotes the polynomial $p(a_1,\ldots,a_{i-1},x_i,a_{i+1},\ldots,a_n)$.
In the first case, $p(\alpha'(\vX))=0$ and in the third case, $p(\alpha'(\vX))\ge0$ for any assignment $\alpha'\in S$.
In the second case, the sign of $p^{*}$ is positive constantly or negative constantly on the whole real axis.
Since $\ell$ is false under $\alpha$, we have $p(\alpha(\vX))\ge0$, that is $p^{*}(a_i)\ge0$.
So, $p^{*}(x_i)>0$ for any $x_i\in\R$, which means $p(\alpha'(\vX))>0$ for any $\alpha'\in S$.
Therefore, no solution to $\ell$ exists in $S$ in the three cases.
That completes the proof.\hfill$\square$
\end{proof}
\vspace{-2mm}

The above theorem shows that if $\sm(x_i,\ell)$ does not exist, then there is no need to search for a solution to $\ell$ along the line $(a_1,\ldots,a_{i-1},\R,a_{i+1},\ldots,a_n)$.
And we can always obtain a solution to $\ell$ after executing a $\sm(x_i,\ell)$ operation.

\vspace{-2mm}
\begin{example}\label{ex:cell_jump}
Assume the current assignment is $\alpha:x_1\mapsto 1,\;x_2\mapsto 1$.
Consider two false atomic polynomial formulas 
$\ell_1:2x_1^2+2x_2^2-1<0$ and
$\ell_2:x_1^8x_2^3 - 4x_1^6 + 6x_1^4x_2 - 4x_1^2 + x_2>0$.
Let $p_1:=\poly(\ell_1)$ and $p_2:=\poly(\ell_2)$.

We first consider $\sm(x_i,\ell_1)$.
For the variable $x_1$, the corresponding univariate polynomial is $p_1(x_1,1)=2x_1^2+1$, and for $x_2$, the corresponding one is $p_1(1,x_2)=2x_2^2+1$.
Both of them have no real roots, and thus 
there exists no $\sm(x_1,\ell_1)$ operation and no $\sm(x_2,\ell_1)$ operation for $\ell_1$. 
Applying Theorem \ref{thm:cell_jump_xi}, we know a solution of $\ell_1$ can only locate in $\R^{2}\setminus (1,\R)\cup(\R,1)$ $($also see Figure \ref{fig:example} $($a$))$.
So, we cannot find a solution of $\ell_1$ through one-step cell-jump from the assignment point $(1,1)$ along the lines $(1,\R)$ and $(\R,1)$. 

Then consider $\sm(x_i,\ell_2)$. 
For the variable $x_1$, the corresponding univariate polynomial is $p_2(x_1,1)=x_1^8 - 4x_1^6 + 6x_1^4 - 4x_1^2 + 1$.
Recall Example \ref{example:about_sample_points}.
There are two positive sample points of $p_2(x_1,1):$ $-\frac{215}{128},\frac{215}{128}$.
And $\frac{215}{128}$ is the closest one to $\alpha(x_1)$.
So, $\sm(x_1,\ell_2)$ assigns $x_1$ to $\frac{215}{128}$.
After executing $\sm(x_1,\ell_2)$, the assignment becomes $\alpha':x_1\mapsto\frac{215}{128},\;x_2\mapsto 1$ which is a solution of $\ell_2$.
For the variable $x_2$, the corresponding polynomial is $p_2(1,x_2)=x_2^3 + 7x_2 - 8$, which has one real root $1$.
A sequence of isolating intervals of $p_2(1,x_2)$ is $(\frac{19}{32},\frac{215}{128})$, and $\frac{215}{128}$ is the only positive sample point.
So, $\sm(x_2,\ell_2)$ assigns $x_2$ to $\frac{215}{128}$, and then the assignment becomes $\alpha'':x_1\mapsto 1,\;x_2\mapsto\frac{215}{128}$ which is another solution of $\ell_2$.
\end{example}
\vspace{-6mm}
\subsection{Cell-Jump Along a Fixed Straight Line}\label{subsec:cell_jump_dir}
\vspace{-1mm}

Given the current assignment $\alpha$ such that $\alpha(\vX)=(a_1,\ldots,a_n)\in\Q^{n}$, a false atomic polynomial formula $\ell$ of the form $p(\vX)>0$ or $p(\vX)<0$ and a vector $dir=(d_1,\ldots,d_n)\in\Q^{n}$, 
we propose Algorithm \ref{alg:cell_jump_dir} to find
a cell-jump operation along the straight line $L$ specified by the point $\alpha(\vX)$ and the direction $dir$, denoted as $\sm(dir,\ell)$.

In order to analyze the values of $p(\vX)$ on line $L$, we introduce a new variable $t$ and replace every $x_i$ in $p(\vX)$ with $a_{i}+d_{i}t$ to get $p^{*}(t)$.
If $\rela(\ell)=$`$<$' and 
$p^{*}(t)$ has negative sample points,
there exists a $\sm(dir,\ell)$ operation. 
Let $t^{*}$ be a negative sample point of $p^{*}(t)$ closest to $0$. 
The assignment becomes $\alpha':x_1\mapsto a_{1}+d_{1}t^{*},\ldots,x_n\mapsto a_{n}+d_{n}t^{*}$ after executing the operation $\sm(dir,\ell)$.
It is obvious that $\alpha'$ is a solution to $\ell$.
If $\rela(\ell)=$`$>$' and $p^{*}(t)$ has positive sample points,
the situation is similar.
Otherwise, $\ell$ has no cell-jump operation along line $L$.
\vspace{-7mm}
\begin{algorithm}[h]
\scriptsize
\DontPrintSemicolon
\LinesNumbered
\setcounter{AlgoLine}{0}
\SetKwInOut{Input}{Input}
\SetKwInOut{Output}{Output}
\Input{
$\alpha=(a_1,\ldots,a_n)$, the current assignment $x_1\mapsto a_1,\ldots,x_n\mapsto a_n$ where $a_i\in\Q$ 

$\ell$, a false atomic polynomial formula under $\alpha$ with a relational operator `$<$' or `$>$'

$dir=(d_1,\ldots,d_n)$, a vector in $\Q^n$
}
\Output{
$\alpha'$, the assignment after executing a $\sm(dir,\ell)$ operation, which is a solution to $\ell$;
FAIL, if there exists no $\sm(dir,\ell)$ operation
}
\caption{{\bf Cell-Jump Along a Fixed Straight Line}}\label{alg:cell_jump_dir}
\BlankLine
$p\leftarrow \poly(\ell)$

$p^{*}\leftarrow$ replace every $x_i$ in $p$ with $a_i+d_i t$, where $t$ is a new variable

\If{$rela(\ell)=$`$<$' and $p^{*}$ has negative sample points}
{
$t^{*}\leftarrow$ a negative sample point of $p^{*}$ closest to $0$

$\alpha'\leftarrow (a_{1}+d_{1}t^{*},\ldots,a_n+d_{n}t^{*})$

\Return{$\alpha'$}
}

\If{$rela(\ell)=$`$>$' and $p^{*}$ has positive sample points}
{
$t^{*}\leftarrow$ a positive sample point of $p^{*}$ closest to $0$

$\alpha'\leftarrow (a_{1}+d_{1}t^{*},\ldots,a_n+d_{n}t^{*})$

\Return{$\alpha'$}
}

\Return{FAIL}
\end{algorithm}
\vspace{-6mm}

Similarly, we have:
\vspace{-2mm}
\begin{theorem}\label{thm:cell_jump_dir}
Suppose the current assignment is $\alpha:x_1\mapsto a_1,\ldots, \allowbreak x_n\mapsto a_n$ where $a_i\in\Q$.
Let $\ell$ be a false atomic polynomial formula under $\alpha$ with a relational operator `$<$' or `$>$', $dir:=(d_1,\ldots,d_n)$ a vector in $\Q^{n}$ and $L:=\{(a_1+d_1t,\ldots,a_n+d_nt)\mid t\in\R\}$.
There exists a solution of $\ell$ in $L$
if and only if there exists a $\sm(dir,\ell)$ operation.
\end{theorem}
\vspace{-1mm}

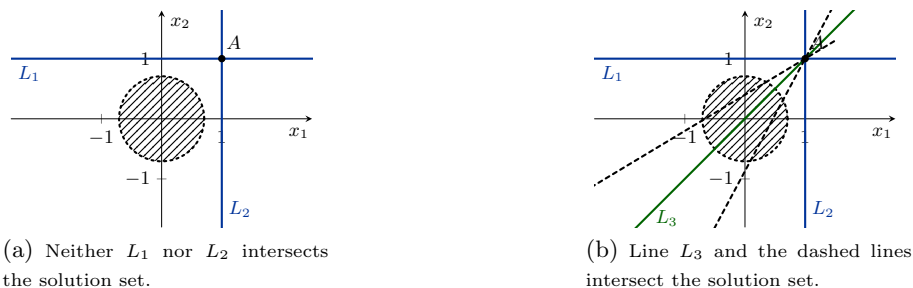
\begin{figure}[h]
\definecolor{qqwuqq}{rgb}{0,0.39215686274509803,0}
\definecolor{qqttzz}{rgb}{0,0.2,0.6}
\definecolor{ududff}{rgb}{0.30196078431372547,0.30196078431372547,1}
\definecolor{uuuuuu}{rgb}{0.26666666666666666,0.26666666666666666,0.26666666666666666}

\subfigure[\scriptsize
Neither $L_1$ nor $L_2$ intersects the solution set.
]
    {
        \centering
        \begin{tikzpicture}[scale=0.8,line cap=round,line join=round,>=triangle 45,x=1cm,y=1cm]
    \begin{axis}[
    x=1cm,y=1cm,
    axis lines=middle,
    xmin=-2.5,
    xmax=2.5,
    ymin=-1.8,
    ymax=1.8,
    xtick={-1,0,1},
    ytick={-1,0,1}]
    \clip(-8.590825495472101,-4.438436597195295) rectangle (9.917963951288046,4.622148812606115);
    \draw [line width=1pt,dotted,fill=black,pattern=north east lines,pattern color=black] (0,0) circle (0.7071067811865475cm);
    \draw [line width=1pt,color=qqttzz,domain=-8.590825495472101:9.917963951288046] plot(\x,{(--3-0*\x)/3});
    \draw [line width=1pt,color=qqttzz] (1,-4.438436597195295) -- (1,4.622148812606115);
    \draw (2,-0.03) node[anchor=north west] {$x_{1}$};
    \draw (0.03,1.8) node[anchor=north west] {$x_{2}$};
    \begin{scriptsize}
    \draw [fill=black] (1,1) circle (1.5pt);
    \draw[color=black] (1.2,1.260768530226581) node {$A$};
    \draw[color=qqttzz] (-2.2,0.75) node {$L_{1}$};
    \draw[color=qqttzz] (1.3,-1.5) node {$L_2$};
    \end{scriptsize}
    \end{axis}
\end{tikzpicture}
    }
\hfill
\subfigure[\scriptsize
Line $L_3$ and the dashed lines intersect the solution set.
]
    {
    \centering
    \begin{tikzpicture}[scale=0.8,line cap=round,line join=round,>=triangle 45,x=1cm,y=1cm]
    \begin{axis}[
    x=1cm,y=1cm,
    axis lines=middle,
    xmin=-2.5,
    xmax=2.5,
    ymin=-1.8,
    ymax=1.8,
    xtick={-1,0,1},
    ytick={-1,0,1}]
    \clip(-8.590825495472101,-4.438436597195295) rectangle (9.917963951288046,4.622148812606115);
    \draw [line width=1pt,dotted,fill=black,pattern=north east lines,pattern color=black] (0,0) circle (0.7071067811865475cm);
    \draw [line width=1pt,color=qqttzz,domain=-8.590825495472101:9.917963951288046] plot(\x,{(--3-0*\x)/3});
    \draw [line width=1pt,color=qqttzz] (1,-4.438436597195295) -- (1,4.622148812606115);
    \draw [line width=1pt,color=qqwuqq,domain=-8.590825495472101:9.917963951288046] plot(\x,{(-0--1*\x)/1});
    \draw (2,-0.03) node[anchor=north west] {$x_{1}$};
    \draw (0.03,1.8) node[anchor=north west] {$x_{2}$};
\draw [line width=1pt,dash pattern=on 2pt off 2pt,domain=-10.387062729306384:1.5311150742190547] plot(\x,{(-0.479848261008447-0.7231001624628146*\x)/-1.2029484234712615});
\draw [line width=1pt,dash pattern=on 2pt off 2pt,domain=-10.387062729306384:1.5311150742190547] plot(\x,{(--0.553364676707671-1.1897124569152553*\x)/-0.6363477802075843});
    \begin{scriptsize}
    \draw [fill=black] (1,1) circle (1.5pt);
    \draw[color=qqttzz] (-2.2,0.75) node {$L_{1}$};
    \draw[color=qqttzz] (1.3,-1.5) node {$L_2$};
    \draw[color=qqwuqq] (-1.3,-1.65) node {$L_3$};
    \draw[color=black] (1.2,1.260768530226581) node {$A$};
    \end{scriptsize}
    \end{axis}
\end{tikzpicture}
    }
\vspace{-4mm}
\caption{
The figure of the cell-jump operations along the lines $L_1$, $L_2$ and $L_3$ for the false atomic polynomial formula $\ell_1:2x_1^2+2x_2^2-1<0$ under the assignment $\alpha:x_1\mapsto1,x_2\mapsto1$.
The dashed circle denotes the circle $2x_1^2+2x_2^2-1=0$ and the shaded part in it represents the solution set of the atom.
The coordinate of point $A$ is $(1,1)$.
Lines $L_1$, $L_2$ and $L_3$ pass through $A$ and are parallel to the $x_1$-axis, the $x_2$-axis and the vector $(1,1)$, respectively.
}
\label{fig:example}
\end{figure}

Theorem \ref{thm:cell_jump_dir} implies that through one-step cell-jump from the point $\alpha(\vX)$ along any line that intersects the solution set of $\ell$, a solution to $\ell$ will be found. 
\vspace{-2mm}
\begin{example}\label{ex:cell_jump_dir} 
Assume the current assignment is $\alpha:x_1\mapsto 1,\;x_2\mapsto 1$.    
Consider the false atomic polynomial formula 
$\ell_1:2x_1^2+2x_2^2-1<0$ in Example \ref{ex:cell_jump}. Let $p:=\poly(\ell_1)$.
By Figure \ref{fig:example} $($b$)$, the line $($line $L_3)$ specified by the point $\alpha(\vX)$ and the direction vector $dir=(1,1)$ intersects the solution set of $\ell_1$.
So, there exists a $\sm(dir,\ell_1)$ operation by Theorem \ref{thm:cell_jump_dir}.
Notice that the line can be described in a parametric form, that is $\{(x_1,x_2)\mid x_1=1+t, x_2=1+t~{\rm where}~t \in\R\}$.
Then, analyzing the values of $p(\vX)$ on the line is equivalent to analyzing those of $p^{*}(t)$ on the real axis, where $p^{*}(t)=p(1+t,1+t)=4t^2 + 8t + 3$.
A sequence of isolating intervals of $p^{*}$ is $(-\frac{215}{128}, -\frac{75}{64})$, $(-\frac{19}{32}, -\frac{61}{128})$, and there are two negative sample points: $-\frac{75}{64}$, $-\frac{19}{32}$.
Since $-\frac{19}{32}$ is the closest one to $0$, the operation $\sm(dir,\ell_1)$ changes the assignment to $\alpha':x_1\mapsto \frac{13}{32},\;x_2\mapsto \frac{13}{32}$, which is a solution of $\ell_1$.
Again by Figure \ref{fig:example}, there are other lines $($the dashed lines$)$ that go through $\alpha(\vX)$ and intersect the solution set.
So, we can also find a solution to $\ell_1$ along these lines.
Actually, for any false atomic polynomial formula with `$<$' or `$>$' that really has solutions, there always exists some direction $dir$ in $\Q^{n}$ such that $\sm(dir,\ell)$ finds one of them.
Therefore, the more directions we try, the greater the probability of finding a solution of $\ell$.
\end{example}
\vspace{-8mm}

\vspace{-8mm}

\begin{remark} 
For a false atomic polynomial formula $\ell$ with `$<$' or `$>$', 
$\sm(x_i,\ell)$ and $\sm(dir,\ell)$ make an assignment move to a new assignment, and both assignments map to an element in $\Q^{n}$.
In fact, we can view $\sm(x_i,\ell)$ as a special case of $\sm(dir,\ell)$ where the $i$-th component of $dir$ is $1$ and all the other components are $0$.
The main difference between 
$\sm(x_i,\ell)$ and $\sm(dir,\ell)$ is that $\sm(x_i,\ell)$ only changes the value of one variable while
$\sm(dir,\ell)$ may change the values of many variables.
The advantage of $\sm(x_i,\ell)$ is to avoid that some atoms can never become true when the values of many variables are adjusted together.
However, performing $\sm(dir,\ell)$ is more efficient in some cases, since it may happen that a solution to $\ell$ can be found through one-step $\sm(dir,\ell)$, but through many steps of $\sm(x_i,\ell)$. 
\end{remark}
\vspace{-1mm}

\vspace{-4mm}
\section{Scoring Functions}\label{sec:score_func}
\vspace{-2mm}
Scoring functions guide local search algorithms to pick an operation at each step.
In this section, we introduce a score function which 
measures the difference of the distances to satisfaction under the assignments before and after performing an operation.



First, we define the distance to truth of an atomic polynomial formula.
\vspace{-1mm}
\begin{definition}[Distance to Truth]\label{def:dtt_atom}
Given the current assignment $\alpha$ such that $\alpha(\vX)=(a_1,\ldots,a_n)\in\Q^{n}$ and a positive parameter $pp\in\Q_{>0}$, 
for an atomic polynomial formula $\ell$ with $p:=\poly(\ell)$, its \emph{distance to truth} 
is
\begin{equation*}
\dtt(\ell,\alpha,pp):=\;\begin{cases}
0, &\text{if}~\alpha~{\text is~a~solution~to}~\ell,\\
|p(a_1,\ldots,a_n)|+pp, &\text{otherwise}.
\end{cases}  
\end{equation*}
\end{definition}
\vspace{-1.5mm}

For an atomic polynomial formula $\ell$,
the parameter $pp$ is introduced to guarantee that the distance to truth of $\ell$ is $0$ if and only if the current assignment $\alpha$ is a solution of $\ell$.
Based on the definition of $\dtt$, we use the method of \cite[Definition 3 and 4]{cai2022local} to define the distance to satisfaction of a polynomial clause and 
the score of an operation, respectively. 

\vspace{-2mm}
\begin{definition}[Distance to Satisfaction]
Given the current assignment $\alpha$ 
and a parameter $pp\in\Q_{>0}$, 
the \emph{distance to satisfaction} of a polynomial clause $c$ is $\dts(c,\alpha,pp):=\min_{\ell\in c}\{\dtt(\ell,\alpha,pp)\}$.
\end{definition}
\vspace{-4mm}




\begin{definition}[Score]
Given a polynomial formula $F$, the current assignment $\alpha$ and a parameter $pp\in\Q_{>0}$,
the \emph{score} of an operation $op$ is defined as
\begin{align*}
\score(op,\alpha,pp):=\;\sum_{c\in F}(\dts(c,\alpha,pp)-\dts(c,\alpha',pp))\cdot w(c),
\end{align*}
where $w(c)$ denotes the weight of clause $c$, and $\alpha'$ is the assignment after performing $op$.
\end{definition}
\vspace{-2mm}

Remark that the definition of the score is associated with the weights of clauses.
In our algorithm, we employ the probabilistic version of
the PAWS scheme \cite{talupur2004range,cai2013local} to update clause weights.
The initial weight of every clause is $1$.
Given a probability $sp$, 
the clause weights are updated as follows:
with probability $1-sp$, the weight of every
falsified clause is increased by one, and with probability $sp$, for every satisfied
clause with weight greater than $1$, the weight is decreased by one.

\vspace{-3mm}
\section{The Main Algorithm}\label{sec:main_alg}
\vspace{-2mm}
Based on the proposed cell-jump operation (see Section \ref{sec:cell_jump}) and scoring function (see Section \ref{sec:score_func}), 
we develop a local search algorithm, called LS Algorithm, for solving satisfiability of polynomial formulas in this section.
The algorithm is a refined extension of the general local search framework as described in Section \ref{sec:general_LS}, where we design a two-level operation selection.
The section also explains the restart mechanism and an optimization strategy used in the algorithm.

Given a polynomial formula $F$ such that every relational operator appearing in it is `$<$' or `$>$' and an initial assignment that maps to an element in $\Q^{n}$,
LS Algorithm (Algorithm \ref{alg:LS}) searches for a solution of $F$ from the initial assignment,
which has the following four steps:
\vspace{-1mm}
\begin{enumerate}


\item\label{item:step1} {\bf Test whether the current assignment is a solution if the terminal condition is not reached.}
If the assignment is a solution, return the solution.
If it is not, go to the next step.
The algorithm terminates at once and returns ``unknown'' if the terminal condition is satisfied.

\item\label{item:step2} {\bf To find a decreasing cell-jump operation along a line parallel to a coordinate axis.} We first need to check that whether such an operation exists.
That is, to determine whether the set $D$ is empty, where
$D=\{\sm(x_i,\ell)\mid \ell~{\rm is~a~false~ atom},\allowbreak
x_i~{\rm appears~in~ \ell}~{\rm and}~\sm(x_i,\ell)~{\rm is~decreasing}\}$.
If $D=\emptyset$, go to the next step.
Otherwise, we adopt the two-level heuristic in \cite[Section 4.2]{cai2022local}.
The heuristic distinguishes a special subset $S\subseteq D$ from the rest of $D$, where $S=\{\sm(x_i,\ell)\in D\mid \ell~{\rm appears~in~a~\allowbreak falsified~clause}\}$,
and searches for an operation with the greatest score from $S$.
If it fails to find any operation from $S$ ({\it i.e.} $S=\emptyset$), then it searches for one with the greatest score from $D\setminus S$.
Perform the found operation and update the assignment. Go to Step $(\ref{item:step1})$.

\item\label{item:step3} {\bf Update clause weights according to the PAWS scheme.}

\item\label{item:step4} {\bf Generate some direction vectors and to find a decreasing cell-jump operation along a line parallel to a generated vector.}
Since it fails to execute a decreasing cell-jump operation along any line parallel to a coordinate axis, we generate some new directions and search for a decreasing cell-jump operation along one of them.
The candidate set of such operations is
$\{\sm(dir,\ell)\mid \ell~{\rm is~a~false~ atom},~dir~{\rm is~a~generated~direction}\allowbreak~
{\rm and}~\sm(dir,\ell)~{\rm is~decreasing}\}.$
If the set is empty, the algorithm returns ``unknown".
Otherwise, we use the two-level heuristic in Step $(\ref{item:step2})$ again to choose an operation from the set.
Perform the chosen operation and update the assignment. Go to Step $(\ref{item:step1})$.
\end{enumerate}
\vspace{-1mm}

We propose a two-level operation selection in LS Algorithm, which prefers to choose an operation changing the values of less variables.
Concretely, only when there does not exist a decreasing $\sm(x_i,\ell)$ operation that changes the value of one variable, do we update clause weights and pick a $\sm(dir,\ell)$ operation that may change values of more variables.
The strategy makes sense in experiments, since changing too many variables together at the beginning might make some atoms never become true.

It remains to explain the restart mechanism and an optimization strategy.
\begin{algorithm}[ht]
\scriptsize
\DontPrintSemicolon
\LinesNumbered
\setcounter{AlgoLine}{0}
\SetKwInOut{Input}{Input}
\SetKwInOut{Output}{Output}
\Input{$F$, a polynomial formula such that the relational operator of every atom is `$<$' or `$>$'

$init_{\alpha}$, an initial assignment that maps to an element in $\Q^{n}$

}
\Output{a solution (in $\Q^{n}$) to $F$ or unknown}
\caption{{\bf LS Algorithm}}\label{alg:LS}
\BlankLine
$\alpha\leftarrow init_{\alpha}$

\While{the terminal condition is not reached}
{
\textbf{if} $\alpha$ satisfies $F$ \textbf{then} \Return $\alpha$

$fal\_cl\leftarrow$ the set of atoms in falsified clauses

$sat\_cl\leftarrow$ the set of false atoms in satisfied clauses

\uIf{$\exists$ a decreasing $\sm(x_i,\ell)$ operation where $\ell\in fal\_cl$\label{algline:op_var_fal}}
{$op\leftarrow$ such an operation with the greatest score

$\alpha\leftarrow\alpha$ with $op$ performed\label{algline:op_var_fal_update}
}
\uElseIf{$\exists$ a decreasing $\sm(x_i,\ell)$ operation where $\ell\in sat\_cl$\label{algline:op_var_sat}}
{$op\leftarrow$ such an operation with the greatest score

$\alpha\leftarrow\alpha$ with $op$ performed\label{algline:op_var_sat_update}
}
\Else{
update clause weights according to the PAWS scheme

generate a direction vector set $dset$

\uIf{$\exists$ a decreasing $\sm(dir,\ell)$ operation where $dir\in dset$ and $\ell\in fal\_cl$\label{algline:op_vec_fal}}
{$op\leftarrow$ such an operation with the greatest score

$\alpha\leftarrow\alpha$ with $op$ performed
}
\uElseIf{$\exists$ a decreasing $\sm(dir,\ell)$ operation where $dir\in dset$ and $\ell\in sat\_cl$\label{algline:op_vec_sat}}
{$op\leftarrow$ such an operation with the greatest score

$\alpha\leftarrow\alpha$ with $op$ performed
}
\Else{\Return{unknown}}
}
}
\Return{unknown}
\end{algorithm}
\noindent\textbf{Restart Mechanism.}
Given any initial assignment, LS Algorithm takes it as the starting point of the local search.
If the algorithm returns ``unknown'', 
we restart LS Algorithm with another initial assignment.
A general local search framework, like Algorithm \ref{alg:LocalSearchProc}, searches for a solution from only one starting point. 
However, the restart mechanism allows us to search from more starting points. 
The approach of combining the restart mechanism and a local search procedure also aids global search, which finds a solution over the entire search space.

\noindent\textbf{Forbidding Strategies.}
An inherent problem of the local search method is cycling, {\it i.e.}, revisiting assignments. 
Cycle phenomenon wastes time and prevents the search from getting out of local minima.
So, we employ a popular forbidding strategy, called tabu strategy \cite{glover1998tabu}, to deal with it.
The tabu strategy forbids reversing the recent changes and can be directly applied in LS Algorithm.
Notice that every cell-jump operation increases or decreases the values of some variables.
After executing an operation that increases/decreases the value of a variable, the tabu strategy forbids decreasing/increasing the value of the variable in the subsequent $tt$ iterations, where $tt\in\Z_{\ge0}$ is a given parameter. 
\vspace{-1mm}

\begin{remark} 
If the input formula has 
equality constraints, then we need to define a cell-jump operation for a false atom of the form $p(\vX)=0$. 
Given the current assignment $\alpha:x_1\mapsto a_1,\ldots, \allowbreak x_n\mapsto a_n~(a_i\in\Q)$, the operation should assign some variable $x_i$ to a real root of $p(a_1,\ldots,a_{i-1},x_i,a_{i+1},\ldots,a_n)$, which may be an algebraic number.
Since it is time-consuming to isolate real roots of a polynomial with algebraic coefficients, we must guarantee that all assignments are rational during the search.
Thus, we restrict that for every equality equation $p(\vX)=0$ in the formula, there exists at least one variable such that the degree of $p$ w.r.t. the variable is $1$.  
Then, LS Algorithm also works for such a polynomial formula after some minor modifications:
In Line \ref{algline:op_var_fal} (or Line \ref{algline:op_var_sat}), for every atom $\ell\in fal\_cl$ (or $\ell\in sat\_cl$) and for every variable $x_i$,
if $\ell$ has the form $p(\vX)=0$, $p$ is linear w.r.t. $x_i$ and $p(a_1,\ldots,a_{i-1},x_i,a_{i+1},\ldots,a_n)$ is not a constant polynomial,
there is a candidate operation that changes the value of $x_i$ to the (rational) solution of $p(a_1,\ldots,a_{i-1},x_i,a_{i+1},\ldots,a_n)=0$; if $\ell$ has the form $p(\vX)>0$ or $p(\vX)<0$,
a candidate operation is $\sm(x_i,\ell)$.
We perform a decreasing candidate operation with the greatest score if such one exists, and 
update $\alpha$ in Line \ref{algline:op_var_fal_update} (or Line \ref{algline:op_var_sat_update}).
In Line \ref{algline:op_vec_fal} (or Line \ref{algline:op_vec_sat}), we only deal with inequality constraints from $fal\_cl$ (or $sat\_cl$), and skip equality constraints.
\end{remark}

\vspace{-4mm}
\section{Experiments}\label{sec:experiment}
\vspace{-1mm}
We carried out experiments to evaluate LS Algorithm on two classes of instances, 
where one class consists of selected instances from SMTLIB while another is generated randomly, and compared our tool with state-of-the-art SMT(NRA) solvers.
Furthermore, we combine our tool with CVC5 to obtain
a sequential portfolio solver, which shows better performance.
\vspace{-3mm}
\subsection{Experiment Preparation}
\vspace{-1mm}
 
 
{\bf Implementation:}
We implemented LS Algorithm with {\tt Maple2022} as a tool, which is also named LS.
There are $3$ parameters in LS Algorithm:
$pp$ for computing the score of an operation,
$tt$ for the tabu strategy and $sp$ for the PAWS scheme, which are set as $pp=1$, $tt=10$ and $sp=0.003$.
The initial assignments for restarts are set as follows:
All variables are assigned with $1$ for the first time.
For the second time, 
for a variable $x_i$,
if there exists clause $x_i< ub\lor x_i=ub$ or $x_i> lb\lor x_i=lb$, then $x_i$ is assigned with $ub$ or $lb$; otherwise, $x_i$ is assigned with $1$.
For the $i$-th time $(3\le i\le7)$, every variable is assigned with $1$ or $-1$ randomly.
For the $i$-th time $(i\ge8)$,
every variable is assigned with a random integer between $-50(i-6)$ and $50(i-6)$.
The direction vectors in LS Algorithm are generated in the following way:
Suppose the current assignment is $x_1\mapsto a_1,\ldots,x_n\mapsto a_n~(a_i\in \Q)$ and 
the polynomial appearing in the atom to deal with is $p$.
We generate $12$ vectors.
The first one is the gradient vector $(\frac{\partial p}{\partial x_1},\ldots,\frac{\partial p}{\partial x_n})|_{(a_1,\ldots,a_n)}$.
The second one is the vector $(a_1,\ldots,a_n)$.
And the rest are random vectors where every component is a random integer between $-1000$ and $1000$.


\noindent{\bf Experiment Setup:}
All experiments were conducted on 16-Core Intel Core i9-12900KF
with 128GB of memory and ARCH LINUX SYSTEM.
We compare our tool with 4 state-of-the-art SMT(NRA) solvers, namely Z3 (4.11.2), CVC5 (1.0.3), Yices2 (2.6.4) and MathSAT5 (5.6.5).
Each solver is executed 
with a cutoff time of $1200$ seconds (as in the SMT Competition) for each instance.
We also combine LS with CVC5 as a sequential portfolio solver, denoted as ``LS+CVC5'', by running LS with a time limit 10 seconds, and then CVC5 from scratch with the remaining 1190 seconds if LS fails to solve the instance.


\vspace{-4mm}
\subsection{Instances}
\vspace{-1.5mm}

We prepare two classes of instances.
One class consists of
totally $2736$ unknown and satisfiable instances from SMTLIB-NRA\footnote{\url{https://clc-gitlab.cs.uiowa.edu:2443/SMT-LIB-benchmarks/QF_NRA}.}, where in every equality polynomial constraint, the degree of the polynomial w.r.t. each variable is less than or equal to $1$.

The rest are random instances.
Before introducing the generation approach of random instances, we first define some notation.
Let $\rn(down,up)$ denote a random integer between two integers $down$ and $up$, and $\rp(\{x_1,\ldots,x_{n}\},d,m)$ denote a random polynomial $\sum_{i=1}^m c_iM_i+c_0$, where $c_i=\rn(-1000,1000)$ for $0\le i\le m$, $M_1$ is a random monomial in $\{x_{1}^{a_1}\cdots x_{n}^{a_{n}}\mid a_i\in\Z_{\ge0},~a_1+\cdots+a_{n}=d\}$ and $M_i~(2\le i\le m)$ is a random monomial in $\{x_{1}^{a_1}\cdots x_{n}^{a_{n}}\mid a_i\in\Z_{\ge0},~a_1+\cdots+a_{n}\le d\}$.

A randomly generated polynomial formula 
$\rf(\{\varnum_{1},\varnum_{2}\},\{\polyum_{1},\polyum_{2}\},\{d_{-},\allowbreak d_{+}\},\{n_{-},n_{+}\},\{m_{-},m_{+}\},
\{\clnum_{1},\clnum_{2}\},\{\cllen_{1},\cllen_{2}\}),
$
where all parameters are in $\Z_{\ge0}$, is constructed as follows:
First, let $n:=\rn(\varnum_1,\varnum_2)$ and generate $n$ variables $x_1,\ldots,x_n$. 
Second, let $num:=\rn(\polyum_1,\polyum_2)$ and generate $num$ polynomials $p_1,\ldots,p_{num}$. 
Every $p_i$ is a random polynomial $\rp(\{x_{i_1},\ldots,x_{i_{n_{i}}}\},d,m)$, where $n_i=\rn(n_{-},n_{+})$,
$d=\rn(d_{-},d_{+})$,
$m=\rn(m_{-},m_{+})$, and
$\{x_{i_1},\ldots,x_{i_{n_i}}\}$ are $n_i$ variables randomly selected from $\{x_1,\ldots,x_n$\}.
Finally, let $\clnum:=\rn(\clnum_1,\allowbreak \clnum_2)$ and generate $\clnum$ clauses such that the number of atoms in a generated clause is $\rn(\cllen_1,\cllen_2)$.
The $\rn(\cllen_1,\cllen_2)$ atoms are randomly picked from $\{p_i<0,p_i>0,p_i=0\mid 1\le i\le num\}$.
If some picked atom has the form $p_i=0$ and there exists a variable such that the degree of $p_i$ w.r.t. the variable is greater than $1$, replace the atom with $p_i<0$ or $p_i>0$ with equal probability.
We generate totally $500$ random polynomial formulas according to
$\rf(\{30,40\},
\{60,80\},
\{20,30\},\{10,20\},\{20,30\},
\{40,60\},
\{3,5\})$.

The two classes of instances have different characteristics.
The instances selected from SMTLIB-NRA usually contain lots of linear constraints, and their complexity is reflected in the propositional abstraction.
For a random instance, all the polynomials in it are nonlinear and of high degrees, while its propositional abstraction is relatively simple. 

\vspace{-4mm}
\subsection{Experimental Results}
\vspace{-1mm}

The experimental results are presented in Table \ref{tab:total_res}.
The column ``\#inst'' records the number of instances.
Let us first see Column ``Z3''--Column ``LS''.
On instances from SMTLIB-NRA, LS performs worse than all competitors except MathSAT5, but it is still comparable.
On random instances, only LS solved all of them, while the competitor Z3 with the best performance solved $29\%$ of them. 
The results show that LS is not good at solving polynomial formulas with complex propositional abstraction and lots of linear constraints, but it has great ability to handle those with high-degree polynomials.
A possible explanation is that as a local search solver, LS cannot exploit the propositional abstraction well to find a solution.
However, for a formula with plenty of high-degree polynomials, cell-jump may `jump' to a solution faster.
The last column shows that LS+CVC5 solved the most instances on both classes of instances.
It means LS and state-of-the-art SMT(NRA) solvers have complementary performance.

\vspace{-4mm}
\begin{table}[ht]\small
    \centering
    \begin{tabular}{cccccccc}
    \hline
                         & \#inst    & Z3   & CVC5 & Yices2 & MathSAT5&  LS & LS+CVC5\\
    \hline
        SMTLIB-NRA       &  2736    & 2519 & {\bf 2563} & 2411 & 1597 & 2246&2601\\
        Random instances &  500      & 145  & 0     & 22 & 0 & {\bf 500} &500 \\
        Total           &  3236      & 2664 & 2563 & 2433 & 1597 & {\bf 2746} & 3101\\
    \hline
    \end{tabular}
    \caption{Results on SMTLIB-NRA and random instances.}
    \label{tab:total_res}
    \vspace{-3.5mm}
\end{table}
\vspace{-6mm}

\begin{figure}[t]
\begin{minipage}[htbp]{0.45\linewidth}
\begin{subfigure}
    \centering
    
    \includegraphics[width=0.9\textwidth]{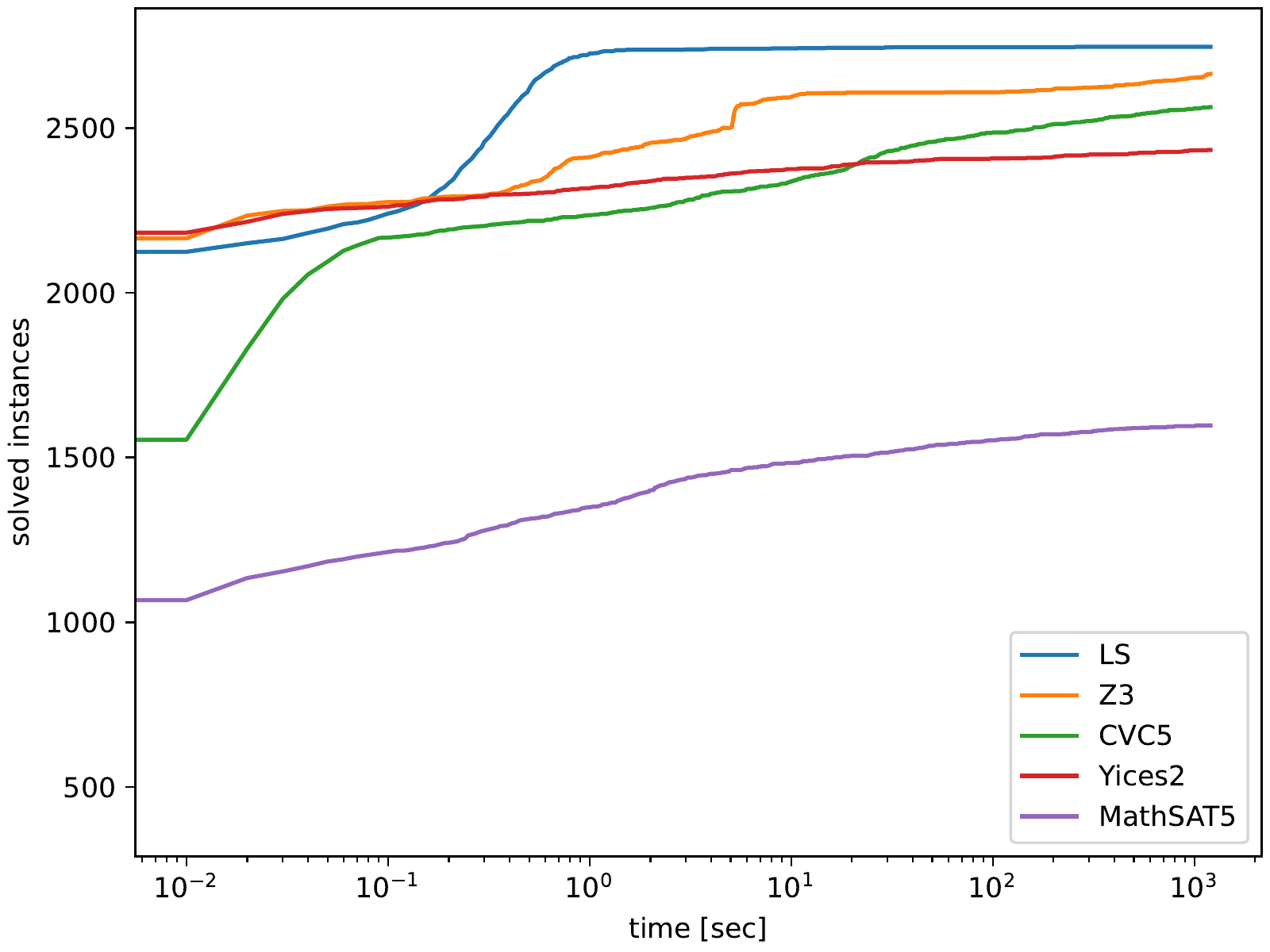}
    \vspace{-3mm}
    \caption{Number of solved instances within given time (sec: seconds).}
    \label{fig:all_compared_line.pdf}
\end{subfigure}
\end{minipage}
\hfill
\begin{minipage}[htbp]{0.45\linewidth}
\begin{subfigure}
    \centering
    \includegraphics[width=0.9\textwidth]{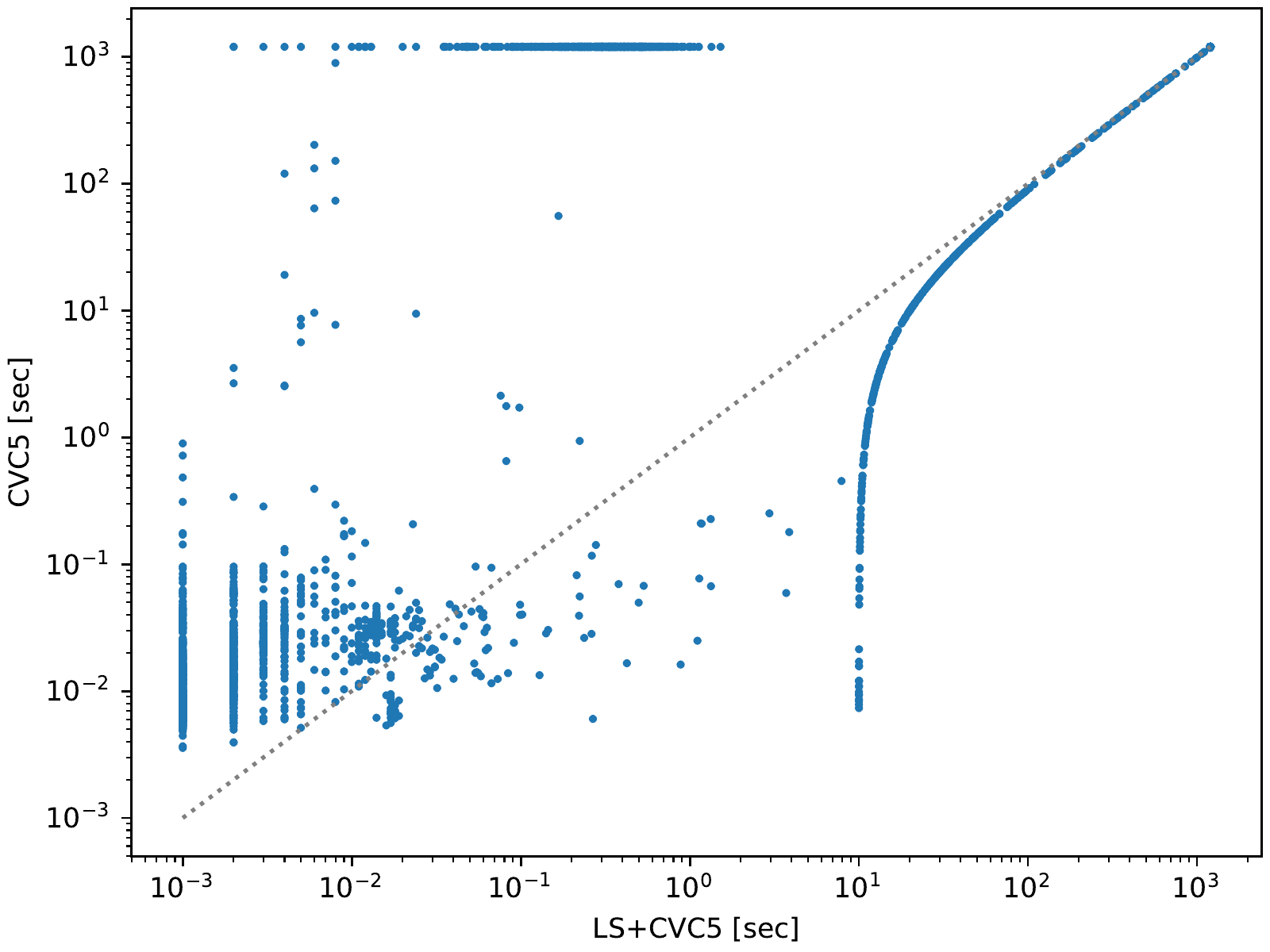}
    \vspace{-5mm}
    \caption{Comparing LS+CVC5 with CVC5.}
    \label{fig:Comparing_maple-cvc5_cvc5}
\end{subfigure}
\end{minipage}
\begin{minipage}[htbp]{0.45\linewidth}
\begin{subfigure}
    \centering
    \includegraphics[width=0.9\textwidth]{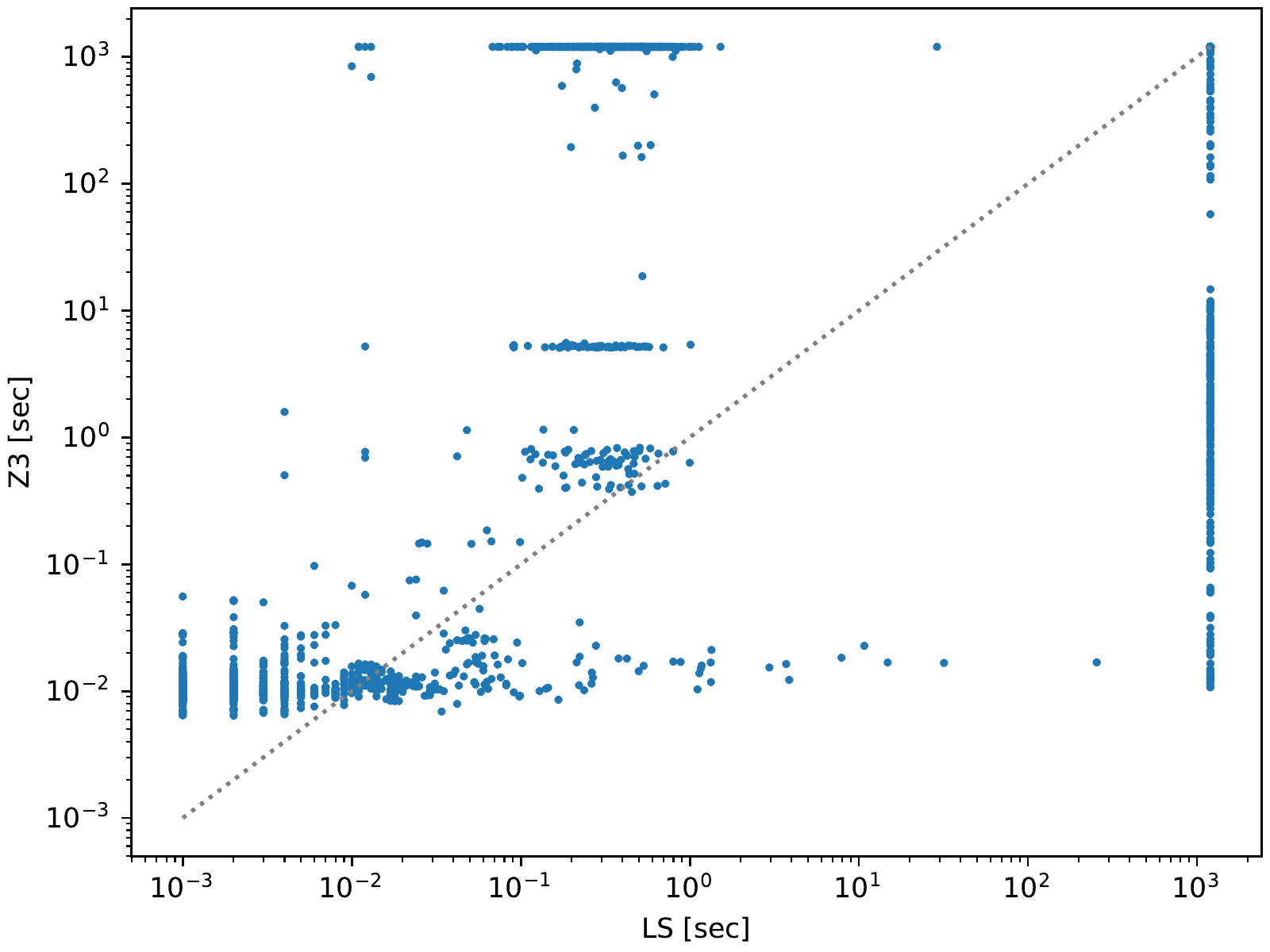}
    \vspace{-5mm}
    \caption{Comparing LS with Z3.}
    \label{fig:Comparing_maple_z3}
\end{subfigure}
\end{minipage}
\hfill
\begin{minipage}[htbp]{0.45\linewidth}
\begin{subfigure}
    \centering
    \includegraphics[width=0.9\textwidth]{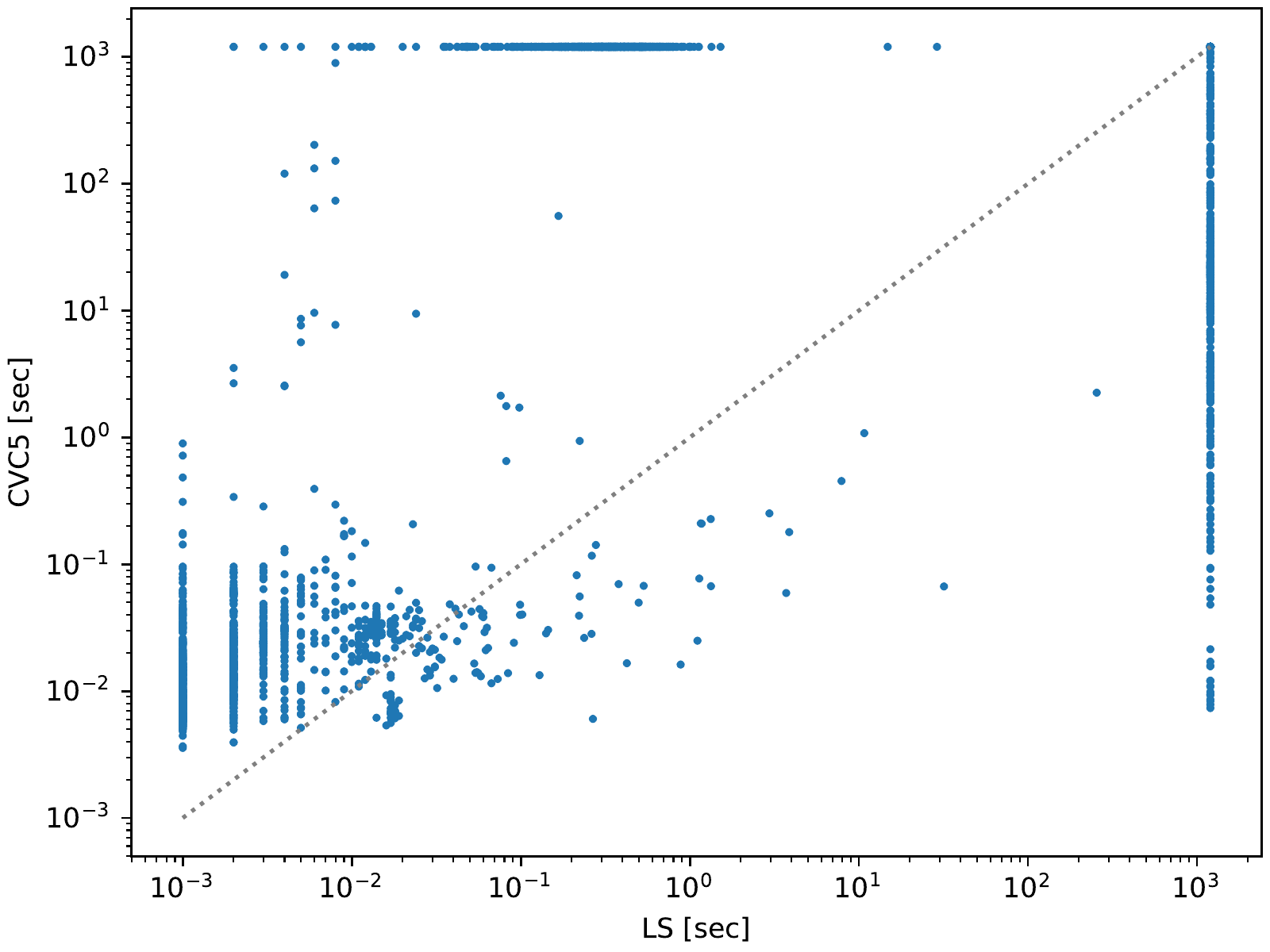}
    \vspace{-5mm}
    \caption{Comparing LS with CVC5.}
    \label{fig:Comparing_maple_cvc5}
\end{subfigure}
\end{minipage}
\begin{minipage}[htbp]{0.45\linewidth}
\begin{subfigure}
    \centering
    \includegraphics[width=0.9\textwidth]{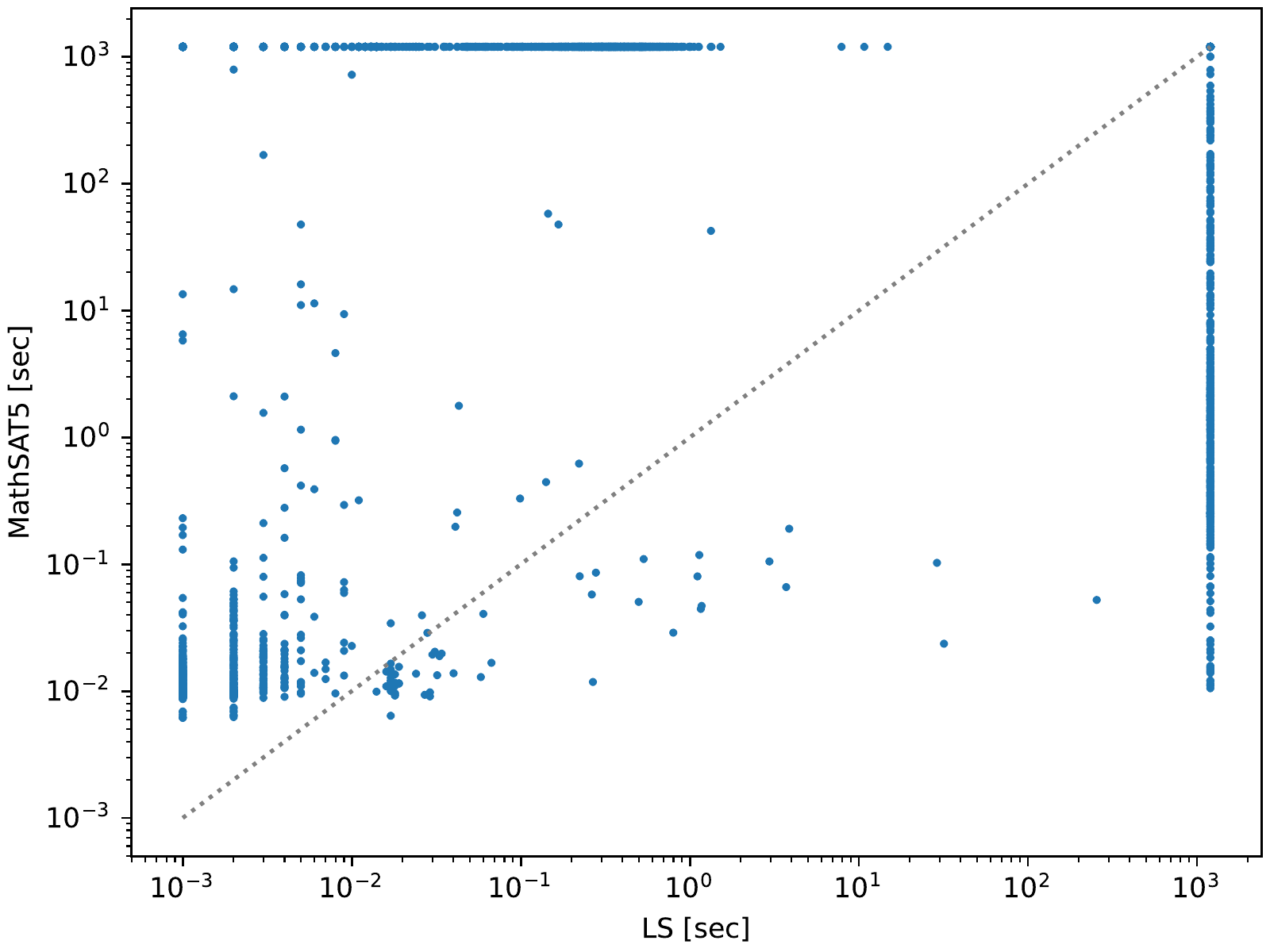}
    \vspace{-5mm}
    \caption{Comparing LS with MathSAT5.}
    \label{fig:Comparing_maple_mathsat}
\end{subfigure}
\end{minipage}
\hfill
\begin{minipage}[htbp]{0.45\linewidth}
\begin{subfigure}
    \centering
    \includegraphics[width=0.9\textwidth]{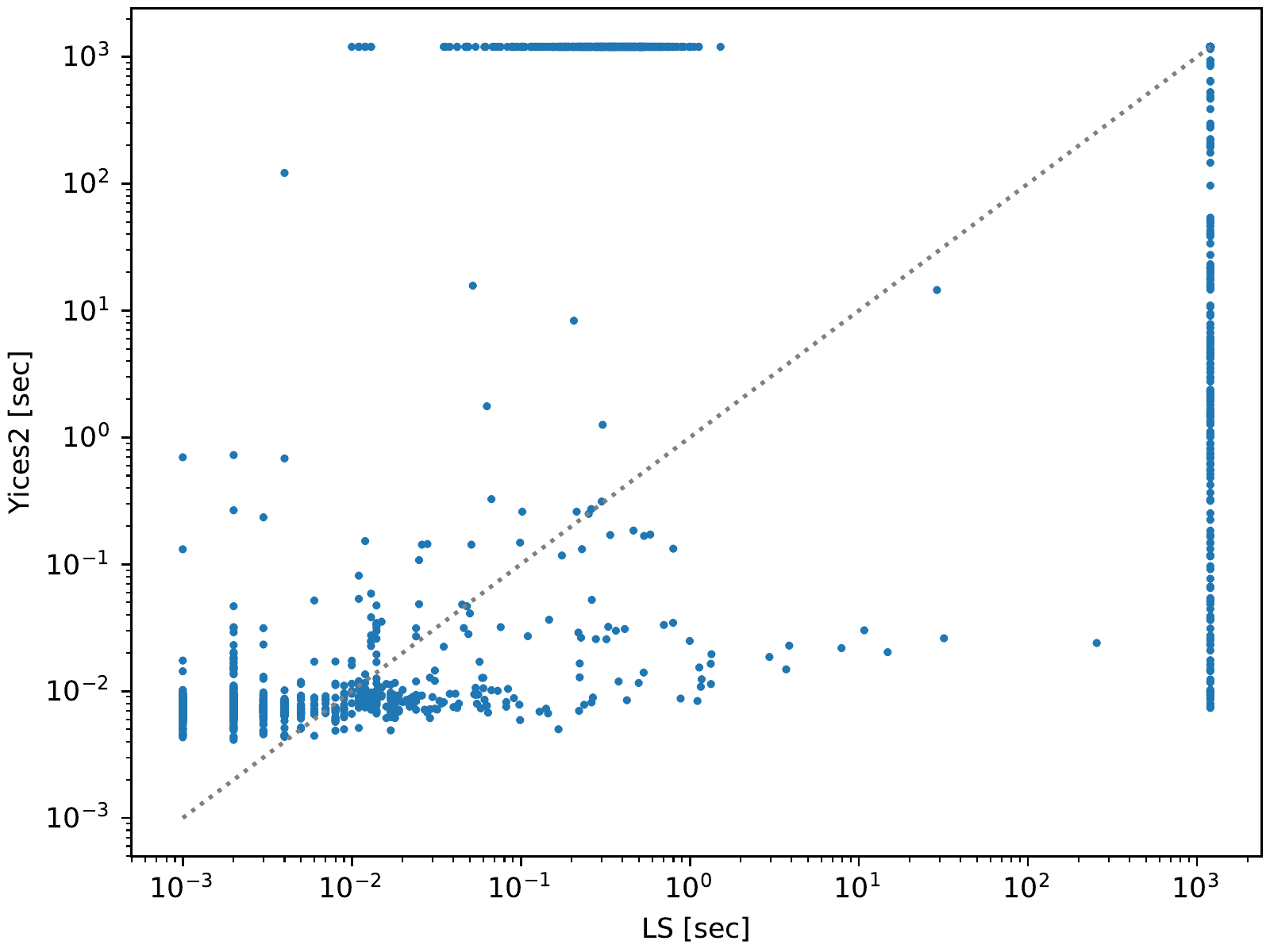}
    \vspace{-5mm}
    \caption{Comparing LS with Yices2.}
    \label{fig:Comparing_maple_yices}
\end{subfigure}
\end{minipage}

\end{figure}

Besides, Figure \ref{fig:all_compared_line.pdf} shows 
the performance of LS and the competitors on all instances.
The horizontal axis represents time, while the vertical axis represents the number of solved instances within the corresponding time.
Figure \ref{fig:Comparing_maple-cvc5_cvc5} presents the run time comparisons between LS+CVC5 and CVC5.
Every point in the figure represents an instance.
The horizontal coordinate of the point is the computing time of LS+CVC5 while the vertical coordinate is the computing time of CVC5 (for every 
instance out of time, we record its computing time as $1200$ seconds).
The figure shows that LS+CVC5 improves the performance of CVC5.
We also present the run time comparisons between LS and each competitor in Figures \ref{fig:Comparing_maple_z3}--\ref{fig:Comparing_maple_yices}. 

\vspace{-2mm}
\section{Conclusion}\label{sec:conclusion}
\vspace{-1mm}
For a given SMT(NRA) formula, although the domain of variables in the formula is infinite, the satisfiability of the formula can be decided through tests on a finite number of samples in the domain. A complete search on such samples is inefficient.
In this paper, we propose a local search algorithm for a special class of SMT(NRA) formulas, where every equality polynomial constraint is linear with respect to at least one variable.
The novelty of our algorithm contains the cell-jump operation
and a two-level operation selection which
guide the algorithm to jump 
from one sample to another heuristically.
The algorithm has been 
applied to lots of benchmarks and the experimental results show that it is competitive with state-of-the-art SMT solvers and is good at solving 
those formulas with high-degree polynomial constraints. 
Tests on a solver developed by combining this local search algorithm with CVC5 indicate that the algorithm is complementary to state-of-the-art SMT(NRA) solvers.
For the future work, we will improve our algorithm such that it is able to handle all polynomial formulas.


\bibliographystyle{splncs04}
\bibliography{ref}

\end{document}